\documentstyle[12pt,aasms4]{article}
\received{January 19, 1999}
\accepted{April 1, 1999}
\slugcomment{To appear in The Astronomical Journal}

\def\msun{M$_{\odot}$}
\def\msuny{M$_{\odot}$yr$^{-1}$}

\begin{document}

\title{The Star Formation History of IZw18}

\author{Alessandra Aloisi\altaffilmark{1}, Monica Tosi\altaffilmark{2} and 
Laura Greggio\altaffilmark{1,2,3}}

\altaffiltext{1}{Dipartimento di Astronomia, Universit\`a di Bologna, 
 Via Ranzani 1, I-40127 Bologna, Italy}
 
\altaffiltext{2}{Osservatorio Astronomico di Bologna,
 Via Ranzani 1, I-40127 Bologna, Italy}
  
\altaffiltext{3}{Universitaets Sternwarte Muenchen, Scheinerstrasse 1, 
 D-81679 Muenchen, Germany}

\authoremail{aloisi@astbo3.bo.astro.it, tosi@astbo3.bo.astro.it, 
 greggio@astbo3.bo.astro.it}

% \author{Alessandra Aloisi}
% \affil{Dipartimento di Astronomia, Universit\`a di Bologna, \\
% Via Zamboni 33, I-40126 Bologna, Italy \\
% e-mail: aloisi@bo.astro.it} 
% \author{Monica Tosi}
% \affil{Osservatorio Astronomico di Bologna, \\
% Via Zamboni 33, I-40126 Bologna, Italy \\
% e-mail: tosi@bo.astro.it}
% \author{Laura Greggio}
% \affil{Dipartimento di Astronomia, Universit\`a di Bologna, \\
% Via Zamboni 33, I-40126 Bologna, Italy \\
% e-mail: greggio@bo.astro.it \\
% \& \\
% Universitaets Sternwarte Muenchen, \\
% Scheinerstrasse 1, D-81679 Muenchen, Germany \\
% e-mail: greggio@usm.uni-muenchen.de}
% \author{Mark Clampin}
% \affil{Space Telescope Science Institute, \\
% 3700 San Martin Drive, Baltimore, MD 21218 \\
% e-mail: clampin@stsci.edu}
% \newpage
% \author{Antonella Nota}
% \affil{Space Telescope Science Institute, \\
% 3700 San Martin Drive, Baltimore, MD 21218 \\
% \& \\
% Affiliated with the Astrophysics Division, 
% Space Science Department of the European Space Agency \\
% e-mail: nota@stsci.edu}
% \author{Marco Sirianni}
% \affil{Dipartimento di Astronomia, Universit\`a di Padova, \\
% Vicolo dell'Osservatorio 5, I-40100 Padova, Italy \\
% e-mail: sirianni@astrpd.pd.astro.it}
% Notice that each of these authors has alternate affiliations, which
% are identified by the \altaffilmark after each name.  The actual alternate
% affiliation information is typeset in footnotes at the bottom of the
% first page, and the text itself is specified in \altaffiltext commands.
% There is a separate \altaffiltext for each alternate affiliation
% indicated above.

\begin{abstract}

The star formation history in IZw18 has been inferred from HST/WFPC2 
archival data. This is done by comparing the derived V, B--V and V, V--I 
color-magnitude 
diagrams and luminosity functions with synthetic ones, based on various sets 
of stellar evolutionary tracks.
At a distance of 10 Mpc, the stars resolved in the 
field of IZw18 allow for a lookback time up to 1 Gyr. 
We find that the main body is not experiencing its first episode of star
formation. Instead, it has been forming
stars over the last 0.5--1 Gyr,  at a rate of 
$\sim$\,1--2\,$\times$\,10$^{-2}$\,M$_{\odot}$yr$^{-1}$kpc$^{-2}$.
A more intense activity of 
6--16\,$\times$\,10$^{-2}$\,M$_{\odot}$yr$^{-1}$kpc$^{-2}$ has
taken place between 15 and 20 Myr ago. For 
the secondary body, the lookback time is 0.2 Gyr at most and the uncertainty 
is much higher, due to the shallower diagrams and the small number of 
resolved stars. The derived range of star formation rate is 
3--10\,$\times$\,10$^{-3}$\,M$_{\odot}$yr$^{-1}$kpc$^{-2}$.

The IMF providing the best fit to the observed stellar populations
in the main body has a slope 1.5, much flatter than in
any similar galaxy analyzed with the same method. In the secondary body,
it is peaked at $\alpha\simeq\,$2.2, closer to Salpeter's
slope ($\alpha$=2.35).

\end{abstract}


\section{Introduction}

In the last twenty years increasing attention has been paid to the study of 
dwarf galaxies in order to understand their crucial role in galaxy formation 
and evolution. In hierarchical clustering theories (White \& Frenk 1991;
Kauffmann, White, \& Guiderdoni 1993) these systems can constitute the 
building blocks from which larger systems have been created by merging, 
while in monolithic collapse scenarios (Tinsley \& Gunn 1976; Tinsley 1980a) 
they have been suggested to represent the debris of massive galaxies unable 
to form stars until z$\sim$1. A population of newly star--forming dwarfs at 
z$<$1 has been also invoked in some evolutionary models (Broadhurst, Ellis, \&
Shanks 1988; 
Babul \& Ferguson 1996; see also Ellis 1997 for a review on the subject)
to reproduce the excess of faint blue galaxies observed in deep photometric 
surveys, the most famous being the Hubble Deep Field (Williams et al. 1996). 

Early--type dwarfs (dEs and dSphs) are gas poor and constituted by 
intermediate and old stellar populations, while late--type dwarfs (dIrrs) 
are gas rich and their light is dominated predominantly by very young stars 
associated with bright HII regions, indicators of an ongoing star formation 
process. The former kind of dwarfs show a smooth luminosity distribution
with a low surface brightness (like Sextants, NGC~147, Leo~I, Fornax, Sculptor, 
etc.), while the latter group appears with a patchy intensity distribution 
(NGC~6822, NGC~1569, IC~1613, etc.) and an intermediate surface brightness 
which can become very high in the bright blue knots of blue compact dwarf
galaxies (hereinafter BCDGs). 

At present it is not well understood if there is an evolutionary 
interconnection between dIrr and dE galaxies
(see e.g. Gallagher 1998), as no consistent picture
of dwarf galaxy evolution has emerged yet. It has been suggested
(Davies \& Phillips 1989; and references therein) that the natural 
evolution of dIrrs could be the condition of dEs through the phase of 
BCDG. In this hypothesis, a strong starburst (intense star formation episode 
concentrated in a very short time) can originate a galactic superwind 
(Heckman 1995) with the mechanical energy supplied by stellar winds and 
supernova explosions generated by newly formed massive stars. This wind 
can blow out all the gas from a dIrr, due to its shallow potential well, 
and transform the galaxy into a gas--poor dE. Indeed, narrowband H$\alpha$ 
images and X-ray maps show evidence of the existence of these superwinds in 
some irregulars and BCDGs, as for example NGC~1569 (Heckman 1995), NGC~1705 
(Meurer et al. 1992), IZw18 (Martin 1996; Petrosian et al. 1997), etc.
On the other hand, taking into account observed chemical, photometric and
kinematic properties of both dwarf irregulars and ellipticals, it seems quite
hard to find an efficient mechanism to transform a late-type dIrr into a dE 
(Jerjen \& Binggeli 1997). The derivation of the star formation history of 
dwarf irregular galaxies and the corresponding identification of objects with 
a starbursting regime become thus of primary importance to gain an insight
into the nature and evolution of dwarf galaxies in general. 

IZw18 (also Mrk 116 or UGCA 166) is possibly the BCDG with the most striking 
properties. At a recession velocity of 745$\,\pm\,$3 km/sec (Dufour, Esteban, 
\& Casta\~{n}eda 1996a), corresponding to a distance of 10 Mpc (H$_0\,$= 75 
km\,sec$^{-1}$Mpc$^{-1}$), this system shows very blue colors. The most 
recent estimates on emission-line corrected broadband images  
give U--B=\,--\,0.88 and B--V=\,--\,0.03 (Van Zee et 
al. 1998). These colors are indicative of a very young 
population, but do not exclude an underlying older one. 
The total mass of IZw18 from the rotation curve at a radius of 
10$^{\prime\prime}\!$--12$^{\prime\prime}$ is estimated  to be
$\sim 10^8$ M$_{\odot}$ (e.g. Davidson \& Kinman 1985; 
Petrosian et al. 1997; Van Zee et al. 1998). 
A large amount of neutral gas is detected all around the system,
totalling $\sim 7\times10^7$ M$_{\odot}$. This  
corresponds to $\sim$70$\%$ of the total mass, but only $ 10^7$
M$_{\odot}$ of HI are associated with the optical part of the galaxy 
(e.g. Lequeux \& 
Viallefond 1980; Van Zee et al. 1998). When discovered by Zwicky (1966), 
IZw18 was described as ``two galaxies separated by 5\farcs6 and 
interconnected by a narrow luminous bridge'', surrounded by two ``very faint 
flares'' at 24$^{\prime\prime}$ northwest. More recent CCD ground--based 
images (Davidson, Kinman, \& Friedman 1989; Dufour \& Hester 1990; hereinafter 
respectively DKF89 and DH90) have revealed a more complex structure: the 
{\it two galaxies} are in fact two 
star--forming regions of the same galaxy (usually indicated as NW and SE 
components), while the {\it two flares} are just the most prominent of a few 
nebulosities surrounding IZw18. These minor systems are roughly aligned 
toward the northwest and were initially believed at the same distance,
but now we know from spectroscopic studies that only one component
(referred to as component C in DKF89) is at the same distance 
as IZw18 
and is physically associated with the main body (Dufour et al. 1996a; 
Petrosian et al. 1997; Van Zee et al. 1998). The 
other diffuse objects have been recognized as background galaxies (see Fig.~1a 
of Dufour et al. 1996b, D96, for an overview of the whole IZw18 system). In 
the following we will refer to IZw18 and to component C respectively as main 
body and secondary body (or companion) of IZw18.
Both systems have been resolved into single stars   
for the first time only with HST/WFPC2 by Hunter \& Thronson (1995, 
hereinafter HT95) and D96.  
Indeed, IZw18 and its companion 
irregular galaxy are currently one of the most distant systems ever 
resolved into stars.

This apparently insignificant BCDG became famous right after its discovery,
when Searle \& Sargent (1972) measured from its emission-line spectrum an 
oxygen abundance [O/H]= --1.14, corresponding to only 7$\%$ of the solar value 
and indicating a quite unprocessed gas content. Furthermore, the first studies 
on its color and composition (Sargent \& Searle 1970; Searle \& Sargent 1972; 
Searle, Sargent, \& Bagnuolo 1973) already emphasized  a current star
formation rate (SFR) much 
higher than the mean value in the past. All these observational evidences 
brought to the formulation of the basic question on the nature of IZw18: is it 
a young galaxy which is presently experiencing its first burst of star 
formation, or is it an old system which has already formed stars in the past 
in at least another episode of star formation ?

Subsequent spectroscopic studies in IZw18 (Lequeux et al. 1979; French 1980; 
Kinman \& Davidson 1981; Davidson \& Kinman 1985; Dufour, Garnett, \& 
Shields 1988; Garnett 1989, 1990; Pagel et al. 1992; Skillman \& Kennicutt 
1993; Kunth et al. 1994; Stasi\'nska \& Leitherer 1996; Garnett et al. 1997; 
Izotov \& Thuan 1998) have confirmed its extreme metal deficiency, around 
1/30--1/50 of Z$_{\odot}$. Despite many efforts to detect other galaxies 
with very low metallicity (Terlevich, Skillman, \& Terlevich 1995), IZw18 still
remains the galaxy with the lowest metal and helium content known so far. This 
makes the system a fundamental point in the derivation of the primordial 
helium abundance (Izotov, Thuan, \& Lipovetsky 1994, 1997; Olive, Steigman, 
\& Skillman 1997; Izotov \& Thuan 1998) and in the study of the properties 
of chemically unevolved galaxies. However, there are several observational 
indications that IZw18 is not a primordial galaxy, for instance the presence 
of relatively high C/O and N/O abundance ratios justified only with an earlier 
population of low and intermediate mass stars (Dufour et al. 1988; Garnett et 
al. 1997) and the photometric evidence of an underlying red stellar population,
both from surface photometry of the whole galaxy in the NIR (Thuan 1983) and 
from photometry of single stars in the optical bands (HT95, D96).

For nearby galaxies the safest determination of their SF 
history is obtained resolving their stellar population into single stars 
and inferring their SFR and initial mass function (IMF) with the synthetic 
color--magnitude diagram (CMD) method. This method was first developed for 
dwarf irregular galaxies in the Local Group observed with ground-based 
telescopes (Ferraro et al. 1989; Tosi et al. 1991, hereinafter TGMF; Greggio et 
al. 1993, GMTF; Marconi et al. 1995, MTGF) and has now been updated (Greggio et 
al. 1998, hereinafter G98) for an optimized application to galaxies observed 
with HST. A procedure for the comparison between observed and synthetic CMDs 
has been developed also by Tolstoy \& Saha (1996), who have introduced the 
concept of Bayesian inference to give the relative likelihood of different 
models to constitute a suitable representation of the data. Gallart et al. 
(1996) also follow a similar approach and have introduced in the method the 
concept of metallicity evolution following a given law for the chemical 
enrichment of the interstellar medium. Here we have applied the synthetic CMD 
method to IZw18 using the HST archive data from HT95 and D96. 

The data reduction is described in Sect.2 and the resulting CMDs and LFs in
Sect.3. The method and a description of the comparison of these
data with theoretical synthetic CMDs and LFs are given in Sect.4, with
the resulting conclusions on the recent evolution of IZw18 and its companion.
An overall discussion of these conclusions in the
framework of the current common knowledge on this galaxy is finally given
in Sect.5.

\section{Observations and Data Reduction}

IZw18 has been observed with different instruments on board of 
HST\footnote{Observations with the NASA/ESA Hubble
Space Telescope, are obtained at the Space Telescope Science Institute,
which is operated by AURA for NASA under contract NAS5-26555}
by different investigators. For our purposes we have retrieved from the 
HST archive and re-reduced all the HST/WFPC2 images available at January 
1st, 1998. 

\subsection{The data}

Two sets of deep exposures were taken in 
November 1994 and a third set of shorter ones in March 1995. 
In the first set of data (PI: Hunter, GO-5309, November 1994) IZw18 was 
centered on the PC CCD, with an effective plate scale of 
0\farcs045 pixel$^{-1}$ and a field of view corresponding to 
36$^{\prime\prime} \times$ 36$^{\prime\prime}$. The target 
was observed in the three broadband filters F336W, F555W and F814W
(similar to the standard ground-based broadbands U, V and I),
and in the two narrowband filters F469N and F656N (sampling the nebular lines 
HeII $\lambda$4686 and H$\alpha$ $\lambda$6563), in order to map
the ionized gas and WR stars. Results from 
this set of data are presented in HT95.

The second set of exposures (PI: Dufour, GO-5434, November 1994) consists 
of deep frames of IZw18 and its companion system on the WF3 CCD, with a 
plate scale and a field of view of 0\farcs1 and 80$^{\prime\prime} 
\times$ 80$^{\prime\prime}$ respectively. The frames are available in
the three broadband 
filters F450W, F555W and F702W (corresponding indicatively to B, V and
R), and in the two narrowband filters F502N and F658N, mapping the 
two nebular lines [OIII] $\lambda$5007 
and [NII] $\lambda$6583. Photometric results from this set of data are 
presented by D96. 

Finally, the third set of images (PI: Dufour, GO-5434, March 1995), are
in the three broadband filters F439W, F555W and F675W (the F439W is a filter 
in the B band region narrower than the F450W; the F675W is a filter in 
the R band region narrower than the F702W). 

A complete summary of all the data available for IZw18 as observed with the 
HST/WFPC2 is presented in Table 1, where we have indicated 
the filter (column 1), the WFPC2 camera where the target was centered on 
(column 2), the principal investigator (column 3),  the epoch of observation 
(column 4), the integration time in seconds for each single exposure (column 
5) and the image root names (column 6).

We actually used only a subset of all the data available on IZw18, 
as indicated in Table 1 by an asterisk near the image name. In fact, we were 
interested in obtaining the color-magnitude diagrams V, B--V and V, V--I for 
IZw18 with the highest possible resolution (i.e. from the PC camera) 
and the color-magnitude diagram V, B--V of its companion on WF3.
The observations in the narrowband filters were used to take into account
the contribution of the ionized gas in the different bands. 

\subsection{Photometric reduction}

We reduced the data applying all the corrections required
to minimize photometric uncertainties and to achieve the highest photometric 
accuracy. All the reductions were performed in the IRAF 
\footnote {IRAF is distributed by the National Optical Astronomy Observatories,
which is operated by the Association of Universities for Research in Astronomy,
Inc., under cooperative agreement with the National Science Foundation.}
environment.

For each single exposure we corrected warm pixels and 
flagged hot pixels in the data quality files using the IRAF/STSDAS task 
{\it warmpix}. To restore the correct relative count numbers 
between pixels at different positions on the CCD,
we performed geometric distorsion and CTE corrections on each single 
frame by applying respectively the correcting image f1k1552bu.r9h available 
from the Archive (see Leitherer 1995 for more details) and the linear-ramp 
image with the appropriate value depending on the background as indicated in 
Holtzman et al. (1995a, b; hereinafter H95a and H95b).

Multiple frames through each filter in each dataset were simultaneously 
co-added and cosmic-ray removed. We also removed where possible the 
contribution of ionized gas from broadband images, proceeding as follows: 
after having adequately smoothed narrowband images to eliminate pointlike 
sources from gas maps, we calculated with SYNPHOT the percentage of 
flux detected in each narrowband filter and subtracted it from each broadband 
image.

The pre-reduction provided the following data useful for our purposes: 
for IZw18 and its companion on the WF3 camera, four deep 
frames in the F555W, F450W, F502N and F658N filters (total integration time 
of 4,600 seconds for each filter); for IZw18 on the PC camera we have two 
images in the F555W filter, the deepest one obtained from Hunter's frames 
(for a total integration time of 6,600 seconds) and the shorter one from 
Dufour's data (1,200 seconds in total). We preferred not to
combine the two final PC F555W frames,
because of a great rotational
displacement of one frame with respect to the other, which would have implied 
a repixelization and data manipulation. 
Also available are: one deep image 
in the F814W band (6,600 seconds), a less deep frame in the F439W band (2,000 
seconds) and a quite deep frame in F656N filter (total integration time of 
4,200 seconds).  In Figs~\ref{VimagePC} and \ref{VimageWF3} 
we show the deepest WFPC2 images in the F555W filter respectively for IZw18 
on the PC camera and for its companion galaxy on the WF3 detector:
both systems are well resolved into single stars. It is also possible to 
distinguish HII regions in the SE part of IZw18 as well as the NW cluster, 
while 2 bright star clusters are evident in the center and in the NW part
of the companion system.

The photometric reduction of the frames was performed using the 
DAOPHOT package in IRAF for PSF-fitting photometry in crowded fields on both 
original and gas-subtracted broadband images. First we applied the automatic
star detection routine {\it daofind} to the deepest F555W frame
(detection threshold at 4$\sigma$ above the local background level), and 
then we performed an accurate inspection by eye of each single detected object 
to reject any feature misinterpreted as star by the routine 
(namely, nuclei of faint galaxies, PSF tendrils, noise spikes, etc.). 
The identification of stars in the other filters was then forced assuming the 
final positions of the stars detected in the F555W deeper image as input 
coordinates for the starting centering and aperture photometry in the new 
frame. We also tried to force 
the photometry from the deepest F555W image to the rotated F439W image, 
taking into account the coordinate transformation, but we obtained 
systematically higher photometric errors, introduced by transformation 
uncertainties. We therefore decided to couple the F439W image with the
shallower but unrotated F555W one.

In spite of the performed centering, forcing the photometry in the second
band leads to some mismatches.
These have been identified by plotting the distance 
between the centers in the two images as function of the 
F555W instrumental magnitude. To overcome this problem, we used the two 
routines {\it daomatch} and {\it daomaster}, kindly made available by P. 
B. Stetson, to finally match the coordinate lists in the two coupled filters.

In order to obtain the best PSF for our frames we 
experimented three different methods: a) we used some well isolated stars 
(3 or 4) in each frame to build the observed PSF, b) we ran the {\it Tiny Tim} 
software (Krist \& Hook 1996) to obtain a theoretical PSF and, c) we made 
use of the new tool of the WFPC2 PSF library to get empirical PSFs. The 
different PSFs considered give quite similar results for the PC camera; 
eventually we preferred to adopt the observed PSF since it takes into
account all technical conditions occurring at the epoch of data 
acquisition (real focus value, thermal breathing, etc.). For the WF3 camera 
theoretical PSFs seem instead to work better, since in this case we have to 
deal with a more dramatic undersampling of the observed/empirical PSFs which 
introduces a higher photometric uncertainty.

Once all the stars were measured with the PSF--fitting, those with a disturbed 
image (as indicated by the two image--peculiarity indices $\chi^2$ and {\it 
sharpness}) were identified and rejected. The index $\chi^2$ gives essentially 
the ratio of the observed pixel--to--pixel scatter in the fitting residuals to 
the expected scatter based on the values of the detector characteristics 
(readout noise and gain). The {\it sharpness} is related to the intrinsic
angular size of the astronomical object. We removed all the objects with 
$\chi^2\!>\,$3 and {\it sharpness} lower than --1 or larger than +1, i.e. 
objects with size smaller than a star (like cosmic rays or image defects) 
or objects too extended (like blends or semi-resolved star clusters, HII 
regions and galaxies). We also checked individually all the stars in critical 
positions in our reference CMDs (i.e. the CMD of stars with photometric error 
smaller than 0.2 mag in both filters), like
very bright blue and red stars (possible blends, unresolved stellar clusters, 
or unidentified cosmic rays), very faint objects (possible 
peaks of noise in the forced filters, or residuals of cosmic rays detections).
Finally, we checked accurately all the red stars, which are particularly 
important to discriminate among different SF histories. 

\subsection{Calibration}

The instrumental magnitudes obtained with the PSF--fitting technique were then 
converted into calibrated magnitudes following the prescription of H95a and 
H95b. Since in H95a and H95b the standard calibration is given for an aperture 
of 0\farcs5, we transformed the instrumental magnitudes on an 
aperture radius of 2 pixels into the corresponding ones on an aperture radius 
of 0\farcs5 (11 pixels for PC and 5 pixels for WF3) by calculating
the aperture correction.

This turned out to be a very delicate step, due to the small number of isolated
stars, suitable for aperture photometry, in the field of
IZw18. The derived aperture corrections strongly depend on
the choice of the adopted stars, and the small number of good
templates leads to a large statistical uncertainty. Figure~\ref{apcorr}
illustrates this point for the deepest V and I frames.
For all the isolated stars in each frame (5--15 objects) we measured the
instrumental magnitudes based on aperture photometry 
with different radii. For each star, the aperture corrections (i.e. 
the difference between the magnitude at a certain aperture radius and
the PSF-fitting magnitude) are shown as open circles for different
values of the aperture radius. The full dots represent the 
correction averaged over all the measured stars, with a
2$\sigma$ rejection algorithm, and the vertical bars show the
corresponding 2$\sigma$ ranges. To these mean 
aperture corrections we added the encircled energy corrections as
indicated by H95b; the obtained values corresponding to the conventional 
radius of 0\farcs5 are indicated by crosses.

As apparent from Fig.~\ref{apcorr}, the uncertainty in the calibration
increases with the aperture radius, with more points deviating 
from the mean value, as it becomes increasingly difficult to find no 
defects in the outer pixels which are progressively included in the 
calculation. However, once averaged and corrected for the encircled
energy, the aperture correction turns out fairly constant. The mean of
the aperture correction values over the
considered range of aperture radii is indicated in Fig.~\ref{apcorr}
by the horizontal thick line, which corresponds to  
corrections of $-0.38$ mag and $-0.54$ mag in the $V$ and $I$ bands
respectively. These values are almost identical to those given by
H95b, which is encouraging. We thus used directly the
H95b corrections for our deepest F555W and F814W frames.
For homogeneity, we extended the use of H95b aperture corrections to
the other frames of IZw18 on the PC camera,  as
well as to the frames of the companion galaxy 
on WF3, where the aperture corrections are much more difficult to determine
empirically.

We finally corrected our measures for the gain factor, the WF3 normalization 
and the contamination effect, where necessary, and applied the zero 
points to scale the photometry into the WFPC2 synthetic system relative to 
Vega (Table 9 of H95a). When the value of one of these correcting parameters 
was not available (as for the F450W filter)  we took it from public images 
of calibration programs (e.g. the aperture correction for F450W on WF3) or 
assumed it from the corresponding value of similar filters (F439W instead of 
F450W).

\subsection{Completeness analysis}

One of the larger uncertainties affecting galaxy photometry is due to
crowding. Thus we carried out an accurate completeness analysis using 
the DAOPHOT routine {\it addstar}.
For each frame we performed a series of tests, by adding each time $\sim$10\%
of the stars detected in each half-magnitude bin on the original 
image. We then performed a new photometric reduction of the frame using
the same procedure applied to the original frame, and considering the 
same rejection criteria for spurious objects. We then checked how many added 
stars were lost either because not detected by the automatic routine or because
recovered with a large mag difference $\Delta$m which makes them migrate 
to another magnitude bin. We eventually estimated the completeness
factors by averaging the results obtained repeating the test
10--20 times on each frame. Due to the uneven distribution of stars on the 
images, fainter objects are more easily recovered in the outer, less crowded 
regions. In order to derive the average completeness factors affecting our CMDs
we constrained the {\it Addstar} routine to put artificial stars in the frame
regions where the galaxy is actually located. 

For every pair of filters used to construct the CMD, we performed 
the completeness analysis independently on each frame. Since we forced the 
stellar detection in the shallower frame, the overall completeness factor is 
actually that corresponding to the shallower filter: in practice, either 
the B or the I frame. Table 2 shows the completeness factors (percentage of
recovered artificial stars) and the 
corresponding uncertainties as a function of magnitude for all the frames 
considered in our photometric analysis. The listed values take also into
account the star selection with photometric error smaller than 0.2 mag 
described in the next section. The completeness factors are averages over 
the whole galaxy. Clearly, incompleteness
can be quite different from one region to the other, depending on crowding. 
For instance in the most crowded zone in the NW cluster, incompleteness
is so dramatic that we recover only $\sim$20\% of the artificial stars already
at the 23 magnitude. HII regions and the rich star clusters are unresolved 
in our images and are therefore to be considered fully incomplete. Although
it could be more appropriate to apply different completeness factors in 
different regions, when computing our theoretical simulations we apply only 
the average completeness factors in Table 2. Indeed the resolved objects are
not numerous enough to allow us to simulate individual different regions.

The completeness tests allowed us also to evaluate the influence of
blending in our photometry and the goodness of the applied rejection criteria 
($\chi^2$, sharpness, error, etc.).
In the deepest V and I images of IZw18 on the PC detector the rejection 
criteria seem to affect more intermediate/faint magnitudes 
(24$\la$V$\la$27 and 23$\la$I$\la$25), while in the shallower B and V frames 
they remove more objects at brighter magnitudes (22$\la$V$\la$24.5 and 
22$\la$B$\la$24). A large fraction of the artificial stars rejected for
$\chi^2\!>\,$3 and {\it sharpness} outside the range from $-$1 to 1
(actually a few objects, usually 
rejected for the $\chi^2$ parameter) were recognizable as blends: we have been
able to actually see the companion star for $\sim$50\% of the objects rejected 
in the B image, for $\sim$65\% in both the V images and $\sim$75\% in the I 
image. For IZw18's secondary body on WF3 there were very few objects discarded,
and they were all blends in both the V and B frames. This result confirms the 
need of applying these criteria in order to remove from the CMDs spurious 
objects due to blends. 

We also looked at the effect of blending on the derived stellar
magnitude, by selecting the artificial stars which were 
recovered with a magnitude brighter than $\Delta$m=0.25 mag with respect to 
the input value. This happened for 4.7\% and 4.0\% of the stars added
in the deeper V and I frames of IZw18; for 2.9\% and 5.2\% in the shallower 
V and B images, and around 3.5\% in both the V and B frames for the companion
system on WF3. 
These values give an estimate of the frequency of cases in
which blending affects the photometry more than our allowed photometric 
error in the CMDs ($\sigma_{DAO}\!<\,$0.2 mag in both filters). 
We conclude that 
our data are affected by blending at an average level of $\sim$4\%,
which we consider negligible for the interpretation of the CMD with
the simulation procedure.

\subsection{Photometric errors}

Figures~\ref{errPC} and \ref{errWF3} show the behavior of the photometric error 
$\sigma_{DAO}$ (estimated by DAOPHOT) as a function of magnitude, respectively 
for IZw18 in each combined PC image and for the companion galaxy 
in each combined WF3 frame. In these plots we considered all the stars fitted 
in both coupled filters: respectively 568 and 321 objects for the V vs V--I 
and V vs B--V diagrams relative to IZw18, and 117 stars for the V vs B--V 
diagram of the companion. 

When the rejection criteria of $\chi^2$ and {\it sharpness} are applied,
most of the points with high $\sigma_{DAO}$ at bright magnitudes disappear,
since they are most probably small unresolved stellar associations, 
HII regions or blends.

For the theoretical interpretation we will restrict our CMD to objects
with photometric error smaller than $\sim$ 0.2 mag. We notice that 
$\sigma_{DAO}$ remains below 
0.2 mag down to m$\sim$26 in each deeper frame of both IZw18 and the companion,
and down to m$\sim$25 in the less deep ones of IZw18. 
However, $\sigma_{DAO}$ may underestimate by $\sim$20\% the 
total photometric error (Stetson \& Harris 1988) and we have therefore 
derived an independent estimate of the latter using the outcome of
the completeness tests, and looking at the amplitude of the 
difference $\Delta$m between the assigned and recovered magnitudes 
of the artificial stars.  It turned out that for the companion of IZw18,
the upper envelope of the distribution of the $\Delta$m 
coincides with that of the observed $\sigma_{DAO}$, indicating that in this 
case the errors are well estimated. For the PC frames with IZw18, instead, 
$\sigma_{DAO}$ underestimates by more than 20\% the actual error, especially 
at the brightest magnitudes. The larger error estimates $\Delta$m have 
therefore been adopted in the CMD simulations described in Sect.4.

\subsection{Comparison with previous photometric analyses}

We have compared the photometric results obtained for IZw18 in the deep 
F555W and F814W images on the PC camera with the results of HT95. The 
comparison of our CMD (see Fig.~\ref{VIcmdPC}) with that published by HT95 
(see their Fig.~5) shows that we reach a fainter limiting magnitude. Our 
diagram contains many more blue and red stars fainter than V$\sim$26. 
We have carefully checked the faintest objects to reject any uncertain
detection with ambiguous shape or profile, but most of 
them remain and are localized in the outer and less crowded part of the 
SE star-forming region where the photometry of fainter objects is easier. 
A possible reason for our fainter limiting magnitude could be our  
forcing the photometry in the shallower filter.

We re-measured all the stars listed in Table 2 of HT95, adopting their x and y 
coordinates, in order to directly compare our mags with theirs: despite 
the same photometric package (DAOPHOT) and the same parameters 
for the photometric conversion of the instrumental magnitudes into the 
calibrated ones (zero points, gain, contamination, etc.),
we found a shift in the zero points of $\sim$0.22 in F555W and $\sim$0.14 
in F814W, in the sense that we are 
brighter and bluer than HT95. This can be seen in Fig.~\ref{photcomp}. 
The difference may arise from the use of gas subtracted images or from
the PSFs and aperture corrections. However, we have verified that there is 
no significant 
difference in the photometric reduction of images with and without gas 
subtraction (except for some peculiar objects), and that there is no shift 
in the zero points resulting from the use of PSFs obtained with different 
techniques (as described in \S~2.2). We thus believe that the offset is
due to the difficulty in estimating the aperture corrections
from the stars in the images of this very problematic field.
As described in \S~2.3, there are too few isolated and reliable stars and the
dependence of the photometric correction on the chosen stars, shown in 
Fig.~\ref{apcorr},
may well lead to offsets of some tenths of mag in the calibration.

We have compared also our CMD in the F450W and F555W filters for IZw18's
companion in the WF3 camera (see Fig.~\ref{BVcmdWF3}) with 
the corresponding diagram published by D96 in their Fig.~3, accounting
for the reddening correction. The 
general aspect of the two distributions is quite similar, despite 
some differences. In our CMD there are less faint objects with unphysical 
color (B--V)$\la-$1, probably thanks to the task for 
masking hot and warm pixels (that could be mistaken as faint stars), which
has become available after the photometric reduction by D96. 
There is also an offset in the zero points of both filters, with our 
distribution being $\sim$0.5 mag brighter than D96 in V and in B.
D96 worked on gas-subtracted and rebinned images in order to construct 
an artificial PSF better sampled than the observed one, while we preferred 
to work on the original frames (but, again, we found no differences at all 
using the gas subtracted images) and to use a PSF simulated 
with the {\it Tiny Tim} software. As for the HT95 data, we 
attribute the photometric offsets to uncertainties in the estimate of 
the aperture corrections, enhanced in this case by the 
undersampling of the WF3 camera. 

For what concerns the CMD in the F439W and F555W filters of IZw18 on the PC 
camera, the archival data  have not been published yet. We can therefore 
compare only indirectly our resulting CMD (see our Fig.~\ref{BVcmdPC}) with 
that derived from the F450W and F555W filters on the WF3 detector and 
plotted in D96's Fig.~2. Both diagrams show a fairly large sequence 
and are very similar to each other. When accounting for the reddening, 
our edge of the blue plume 
(at V$\,\simeq\,$22.5), the faint limiting magnitude (at V$\,\simeq\,$27), 
and the average color of the star distribution around (B--V)$\,\simeq\,$0
seem consistent with those in D96.

\section{Observed color-magnitude diagrams and luminosity functions}

\subsection{IZw18 Main Body}

Figures~\ref{VIcmdPC} and \ref{BVcmdPC} show the CMDs derived from the
PC images of IZw18 in the HST
F555W, F814W and F439W bands, that in the following will be referred 
to as the V vs V--I  and the V vs B--V CMDs respectively. 
In panel a) of each figure we plotted 
all the objects measured in both filters with a $\chi^2\!<\,$3 and 
--1$<\!sharpness\!<$1 (respectively 444 in the V vs  V--I and 267 in the V 
vs B--V), while in panel b) we show only the stars with a photometric 
error $\sigma_{DAO}\!<\,$0.2 in both bands, after removing 
spurious detections (247 objects remain in the red CMD and 106 in 
the blue one). The main features 
of the complete diagram remain unaltered after the $\sigma_{DAO}\!<\,$0.2 
selection. Thus we used the CMDs in panels b) as reference diagrams
for the theoretical simulations, as they have a more reliable photometry.

Since this target is located at high galactic latitude (b=+45$^{\circ}$), its 
CMDs do not suffer from significant contamination from foreground stars 
belonging to our Galaxy, and it is not necessary to consider this factor of 
uncertainty in our simulations or to correct for it our observed CMDs. 

A first analysis of the observed V vs V--I diagram of IZw18 (panel b) of 
Fig.~\ref{VIcmdPC}) shows that we 
reached a limiting magnitude of V$\,\simeq\,$26.5 which goes down to 
V$\,\simeq\,$27 for objects with the reddest V--I color. From a morphological 
viewpoint, in this CMD we can easily distinguish the typical blue plume 
observed in ground-based observations of Local Group irregulars: this plume 
is populated both by main-sequence (MS) stars and by stars at the hot edge of 
the post-MS evolutionary phases. For IZw18 the plume extends up to 
V$\,\simeq\,$22 and has a median color V--I$\,\simeq\,$0, indicative of a very 
low reddening. This is in agreement with the most recent values proposed for 
this parameter, e.g. E(B--V)=0.04 in HT95. 
We can also notice the presence in the CMD of several bright supergiants 
with a wide spread of colors from blue to red, 
and some faint red stars. As described in the previous sections, they 
all turned out to be real stars after a detailed analysis of their 
shape and profile. 
All the bright blue and red stars are concentrated in the innermost and more 
crowded regions, particularly in the NW component of IZw18. 

All the HII regions recognized by HT95 (as well as many stars in the badly 
resolved NW cluster), were automatically removed from our V, V--I diagram 
because they turned out to have $\chi^2\!>\,$3 and/or $sharpness$ outside the 
range [--1,1]\,. Also the bright star clusters and associations are rejected
for the same reason. 
This implies that our CMD is not sampling extremely young stars. The 
percentage of flux discarded with this procedure corresponds roughly to 
3.5\% (2\%) of the total flux of the whole galaxy in the V (I) filter, while
it is $\sim$40\% of the total flux sampled by the resolved stars in both bands.

An inspection of the V vs B--V diagram shows that also in this case we reached 
the limiting magnitude V$\,\simeq\,$26. We have however a shallower cut-off as 
the B--V color becomes redder: this is due to the shorter integration time and 
to the lower sensitivity of the F439W filter with respect to the F814W. Again 
in this CMD we can recognize the typical blue plume of the MS and post-MS 
stars, with a median color of B--V$\,\simeq\,$0, and an upper
brightness limit of V$\,\simeq\,$22.5. Also in this case we are not retaining 
the HII regions of HT95 and the star clusters, 
the total flux of the galaxy lost with the rejection criteria being 
1\% (0.5\%) in the V (B) band, again $\sim$40\% of the light in the
resolved stars.

To estimate the masses of the stars visible in the CMDs and the 
corresponding lookback times, we have converted stellar evolutionary tracks 
into the observational plane, and superimposed them on the observed CMDs. 
In Fig.~\ref{tracksPC} we show the Padova tracks with Z=0.0004 
(Fagotto et al. 1994) converted to the V vs V--I (panel a) and V vs
B--V (panel b) plane, having adopted a distance modulus of (m--M)$_0$=
30 and a reddening value E(B--V)=0.04. The V vs V--I diagram 
shows that in IZw18 we have detected MS stars 
with masses higher than 12 M$_{\odot}$ (corresponding to lifetimes younger 
than $\sim$ 20 Myr) and blue-loop stars
with masses down to $\sim$3--4 M$_{\odot}$ (thus with ages up to 
$\sim$0.2 Gyr). The faintest clump of red objects in Fig.~\ref{tracksPC} 
can be populated by (red) core helium burning, asymptotic giant branch (AGB) 
stars and bright red giant branch (RGB) stars, 
whose masses can be in principle as low as $\sim\,$1 \msun,
extending the lookback time up to several Gyr. However, given the large 
photometric error, it is not possible to estimate precisely the mass, and 
therefore the age, of these faint stars. In the following we will 
conservatively assume a lookback time of 1 Gyr.   
In the V vs B--V diagram we see objects with mass larger than 12 M$_{\odot}$ 
in the MS stage ($\tau\la\,$20 Myr) and larger than 7 M$_{\odot}$ in 
the post-MS phase, thus with ages less than $\sim$ 50 Myr. We will use the 
V vs V--I diagram to infer the SF history of IZw18 over the last 
$\sim$ 1 Gyr, while the V vs B--V diagram will be used as a further
check over the last $\sim$ 50 Myr.

In Fig.~\ref{PCLFs} we plot the differential luminosity functions (LFs) of
all the stars with $\sigma_{DAO}\!<\,$0.2 in both filters present in
the V vs V--I and the V vs B--V diagrams (panels a) and b) respectively). 
We can see in both cases a rather smooth trend. The LFs will be used in the 
simulations to check the consistency between models and observations.

It is clear from Fig.~\ref{tracksPC} that the blue plume of IZw18 is
populated by stars both on the MS and at the hot edge of the blue loop 
evolutionary phase and that no safe criterion can be found to separate
the two different populations. For this reason we do not even attempt to
derive a MS-LF, which would be inevitably affected by too large uncertainties
to be of any use.

Also the derivation of the slope of the LF may turn out too uncertain, once
we consider that, as listed in Table 2, the data start to be 
incomplete already at the brightest mags and significantly incomplete at
V=24. For mere sake of comparison with other galaxies, and warning that these
values should only be taken as indicative, we have nonetheless
computed the slope by means of a maximum likelihood fitting on the deeper 
V data. Down to V=23,
where the data are almost complete, but only very few stars are present,
 $\Delta$log\,N/$\Delta$V = 1.28 $\pm$ 0.04; at V=24, where completeness
is 85\%, the slope is 0.68 $\pm 0.02$, and at V=25 (75\% of completeness)
it is 0.45 $\pm$ 0.04. 
The latter value is consistent with those derived by HT95 from the same
data for stars in three different locations (slopes between 0.58 and 0.65),
once we consider that they have corrected them for incompleteness. 

HT95 pointed out that these slopes are steeper than those derived for other
star forming systems like R136 and NGC~604. They also appear steeper
than the average $0.70\pm0.03$ derived by Freedman (1985) and 
Hoessel (1986) from a large sample of irregulars and than those derived
by us in Local Group irregulars (TGMF, GMTF, MTGF)
and in NGC~1569 (G98), since those were derived
in the complete portion of the stellar sample distribution. 
The difference is more striking if one considers that 
the literature values are supposed to refer to MS stars, whereas here we have 
all kind of objects, and in general MS-LFs are steeper than global ones, 
which can include bright supergiants.

As discussed by HT95, it seems unlikely that such steep LF is due
to a steep IMF and we certainly endorse their opinion, since we will show
in the following that the data of IZw18 are actually best reproduced by
assuming a flat and not a steep IMF. We are rather inclined to attribute this
unusual steepness to the particular 
star formation history of the galaxy which can have superimposed around 
V=24--25 two distinct stellar populations, making that mag bin much 
more populated that the brighter ones.

\subsection{IZw18 Companion System}

Figure~\ref{BVcmdWF3} shows the V vs B--V diagram for IZw18's companion, 
resolved into individual stars in the field of view of the WF3 camera. In 
panel a) we plotted the 109 stars measured in both the F555W and F450W filters
with $\chi^2\!<\,$3 and --1$<\!sharpness\!<$1, while in panel b) we considered 
only the subsample of stars (58 objects) with $\sigma_{DAO}\!<\,$0.2, after an 
accurate check for spurious detections. 
It is worth to point out that the rejection criteria eliminate $\sim$1\% 
of the total flux of the secondary body in both B and V filters, and contrary
to IZw18, this corresponds only to $\sim$15\% of all the light in the
measured stars. 
Furthermore the error constraint implies the loss of a lot of 
faint stars, and saves only the brightest part of the blue plume and a 
few red supergiants and giants. The blue plume has a median color 
B--V$\,\simeq\,$0 and reaches the bright limit of V$\,\simeq\,$24 (the 
brightest point at V$\,\simeq\,$22 in panel a) corresponds to the star 
cluster in the center of the system). Given the high galactic latitude of the
system, again we expect no contamination problems. The foreground reddening 
is assumed to be E(B--V)=0.04, as for the main body.

In Fig.~\ref{WF3LF} we report the differential LF in the V band which refers 
to the CMD of the secondary body. The derivation of the slope of the LF is 
even more uncertain than for the main body; we have nonetheless computed it 
by means of a maximum likelihood fitting. Down to V=25, incompleteness is 
low, but only 10 stars are present, and the slope is 
$\Delta$log\,N/$\Delta$V = 0.60 $\pm$ 0.16; down to V=26, where completeness 
is 80\%, the slope is $0.50\pm0.08$. These values are totally consistent with
those derived for other irregulars (see previous subsection).

As already done for the main body, in Fig.~\ref{tracksWF3} we 
show the comparison between the Padova tracks with Z=0.0004 and the V vs B--V 
diagram of the companion galaxy. 
Taking into account the photometric error, the resolved stars could all
be MS objects, yielding a lookback time of only a few tens of Myr at most.
The two brightest red stars, which appear as evolved objects of $\sim$ 7 \msun,
are in fact the two objects circled in Fig.~\ref{VimageWF3}, located rather
far from the bulk of the system, and might therefore be foreground objects.
As an alternative interpretation, the observed CMD can be populated by only a 
few MS stars, with most of the detected objects in the blue loop 
phase. In this case most of the MS would be fainter than our limiting 
magnitude, and the CMD would be sampling the evolved progeny 
of 4 to 9 M$_{\odot}$ objects, yielding a lookback time of $\sim$0.2 Gyr.

From a comparison between the CMDs of IZw18 and its companion it is evident 
that the blue plume of the main system is $\sim\,$2 magnitudes brighter in
V than that of the secondary body: at first glance this may be interpreted 
as an indication of a considerably younger population in the bigger
system. However, the interpretation can be quite
different, once we take into proper account the different contributions
to the blue plume of MS and post-MS stars.

\section{Comparison with synthetic diagrams}

In order to derive the SF history and IMF of IZw18 and its companion, we
have compared the observed CMDs and LFs with theoretical simulations based on
homogeneous sets of stellar evolutionary tracks. The procedure applied here 
for the creation of synthetic CMDs is the same described in detail in TGMF
for ground-based observations of Local Group irregulars 
and in G98
for HST optical data of the nearest starburst dwarf NGC~1569. In the
latter paper a detailed description of the whole procedure and of the 
conversion of synthetic CMDs to the HST/WFPC2 Vega-mag system is also given.

Literature values for the distance to IZw18 range from 9.8 to 11.2
Mpc, corresponding to true distance moduli between 30 and 30.3
mag. For our simulations we adopted (m-M)$_0$ = 30, 
and for the reddening 
E(B--V)=0.04 (HT95), which turned out to provide synthetic MSs 
with average color in agreement with the observed one.

For an adopted IMF, SFR and set of stellar evolutionary 
tracks of a given metallicity, the final product of the simulation is a 
synthetic diagram containing the same number of objects (247 for V, V--I and
106 for V, B--V in the main body, and 58 in the companion) above the same 
limiting magnitude as the observed CMD and 
with the same properties of photometric uncertainties and incompleteness
(Figs \ref{errPC} and \ref{errWF3}, and Table 2).
The free parameters for the CMD simulations are: IMF slope; starting epoch,
duration, and ending epoch of the star formation activity; mode of the
SF (continuous or episodic, constant or exponentially decreasing with time).
For any adopted set of stellar evolution models, we have first generated
synthetic CMDs assuming constant SF throughout all the observable lookback
time and Salpeter's (1955) IMF ($\alpha$=2.35), and
then modified the assumptions on each parameter to see the resulting
effect on the comparison between the predicted CMD and LF 
and the empirical data.
Here we show only a few illustrative cases; for a larger compilation of
model samples see Aloisi \footnote{The Ph.D. Thesis is available in 
electronic form upon request to the first author} (1998, hereinafter A98).

The comparison between the simulated CMDs and the observed one is carryed out
in terms of the major features of the stellar distribution 
in the color-magnitude plane, as for example the relative number of stars in
different evolutionary phases, the color and magnitude of the brightest stars 
or of the blue plume, etc. 
A more quantitative comparison is performed on the LFs. 
We do not perform quantitative tests on the color distribution because 
of the large intrinsic uncertainties in the effective temperatures of 
models of massive stars in their post-MS stage, which are reflected 
in the color determination. 

To constrain the model selection as much as possible, we have simulated 
independently the V, V--I and the V, B--V diagrams and compared the 
corresponding results only a posteriori. This approach may appear 
timewasting, since the two diagrams correspond to the same galactic area 
and therefore represent the same stellar population, but it adds 
independent and useful constraints because stars of different temperature
have quite different weight on the distribution of B--V and V--I CMDs,
and because the photometric errors and incompleteness factors are different
in different frames. Besides, these independent simulations
provide a useful test on the self consistency of the method. The B--V CMD
is in general too shallow to provide by itself reliable information on the SF 
history of IZw18; nonetheless it is very useful for a further selection of 
the models providing the better agreement for the V--I data, thanks to its
higher sensitivity to the younger (i.e. bluer) population. We have
found that only a few of the models selected in
V--I turn out to reproduce also the B--V observed features. 
This has significantly constrained the overall scenario 
able to fit both the B--V and the V--I distributions.

To evaluate the theoretical uncertainties due to different stellar codes, 
input physics, and metallicity effects, it is always important to generate
synthetic CMDs with more than one set of homogeneous tracks, when 
available. Since the overall metallicity of IZw18 and its companion is 
estimated to be between Z$\sim$0.0006 and Z$\sim$0.0004, we have performed our 
simulations for both systems using the Padova tracks with Z=0.0004 (Fagotto et 
al. 1994) and the Geneva tracks with Z=0.001 (Schaller et al. 1992) 
and Z=0.0004 (kindly made available by D. Schaerer). 

These sets of stellar models differ from each other in several aspects, some
of which have significant effects on the synthetic CMDs:
\begin{itemize}
\item At the same nominal metallicity (Z=0.0004) the Padova tracks have the 
same temperatures as the Geneva ones for MS stars (core H-burning phases) and 
for stars at the hot edges of the blue loops (core He-burning phases), but 
lower temperatures for red giants and supergiants, thus spanning a larger 
color range. 
\item The Geneva tracks with Z=0.001 assign lower temperatures to the red 
stars (due to the higher metallicity) and cover the maximum color 
interval, in spite of their slightly cooler hot edges of the 
blue loops.
\item The lifetimes of massive stars at the blue loop edge are systematically 
longer in the Geneva tracks than in the Padova ones. This implies that the 
Geneva models tend to predict post-MS massive stars mostly at the blue edge 
of the loops, whereas the Padova models populate more homogeneously all the 
colors from the red to the blue edges. The opposite occurs for intermediate 
mass stars, for which the Padova models predict longer lifetimes at the blue 
edge than the Geneva ones.
\end{itemize}

\subsection{Main body: Simulations with Padova tracks.}

We have performed about 300 simulations with the set of Padova models with
metallicity Z=0.0004,
testing various values for the slope of the IMF, the starting epoch, the mode 
and the duration of the SF activity, under
the hypotheses of one or two SF episodes occurred during the last 1 Gyr.
All the synthetic CMDs based on these tracks show a short color extension: 
the bluest stars are properly reproduced, but the coolest objects predicted 
by these models have systematically V--I$\leq$1.7, whereas the empirical 
ones are as red as V--I$\,\simeq\,$2. The cause of this inconsistency could 
be an excessively low metallicity parameter in the tracks, or an inadequacy 
of stellar models to reproduce the effective temperatures 
in the coolest phases. Besides, the color conversions are more uncertain
in these phases, due to the much more difficult treatment of molecules in
model atmospheres. On the other hand, 
the calibration of HST data may still be slightly uncertain
and the data may present color equations not properly taken into account.

Independently of the IMF and SFR, the most evident result which clearly 
emerges from the comparison of synthetic and observed CMDs is that we can 
safely exclude that only one single recent burst has occurred, started later 
than a few 10$^7$ years ago. 
This scenario definitely does not allow us to reproduce 
the observed red and blue stars fainter than V$\sim$26 populating the V vs 
V--I diagram, and leads in general to an overabundance of bright blue stars
compared to those observed both in the V--I and in the B--V CMDs.
As an example, in the left hand panels of 
Fig.~\ref{sim1fin} we have plotted the V--I and B--V CMDs 
and corresponding LFs of a SF episode started 
10 Myr ago and still ongoing at a constant rate. 
In this case the IMF is 
steep ($\alpha$=3.0, much steeper than Salpeter's 2.35).
With such a late start, all the objects with masses lower than $\sim$
20 \msun ~are still on the MS and there is no chance to populate the
blue loops at intermediate and faint magnitudes. As a result, the synthetic 
stars are all bluer than V--I$\,\simeq\,$0, at variance with the 
observational distribution. In addition, the 
synthetic LF turns out underpopulated in
the V$\,\ga\,$25.5 portion, and overpopulated in the range V$\,\la\,$23.
Flattening the IMF clearly worsens the result. Also the V, B--V synthetic 
diagram is inconsistent with the data, being populated only with stars
bluer than B--V$\,\simeq\,$0.2.
To fit the data we need an earlier start of the SF activity:
from many tests, we have found that to obtain acceptable results, the SF 
in IZw18 must have started at least 200 Myr ago (see A98).

A continuous SF provides results consistent with the data 
either with a currently ongoing SF, or with one stopped 
not earlier than 5 Myr ago. In the central panels of Fig.~\ref{sim1fin}  
we present the V vs V--I, V vs B--V and relative LFs 
for the case of one episode started 1 Gyr ago with an exponentially 
decreasing SF activity (e-folding time $\tau$=500 Myr) and still ongoing. 
The adopted IMF is $\alpha$=1.5. Notice that in the LF the maximum deviation 
of the model from the observational points is around 3\,$\sigma$. Similar 
results are obtained with a later onset of the SF. With this type of models 
we reproduce fairly well the 
CMDs and LFs, provided that the adopted IMF is flat (1.5$\leq\alpha\leq$1.8). 
Steeper slopes (even when coupled with a constant SFR) lead to worse 
results, since they don't provide enough bright stars when the faint end
of the LF is matched. For instance the best model obtained with a Salpeter's
slope ($\alpha$\,=\,2.35) leads to a LF which deviates in several magnitude
bins by more than 4\,$\sigma$ from the observational points (A98).

Trying to better reproduce the observed color distributions of the stars,
we have considered a two-episode scenario. Models in which the old episode 
occurs from $\sim$1--0.2 Gyr to 100--50 Myr ago, and the young one from 
$\sim$100--30 Myr ago on, provide acceptable results (similar to those 
shown in the central panels of Fig.~\ref{sim1fin} for a single episode) 
when the IMF is flat (1.5$\leq\alpha\leq$1.8). In these cases the predicted 
LF is always within 1--2\,$\sigma$ from the empirical one.
From these simulations we find that the SFR in 
the two episodes is quite similar, and that a significant quiescent 
intermediate phase is not necessary (see A98). At the end of the two-episode
simulations, we can thus assert that IZw18 has experienced a rather continuous 
star-formation activity over a large fraction of the whole lookback time 
sampled by our CMDs.  

The distribution of the yellow and red supergiants in the observed CMD
is quite peculiar: there is a clump of faint red stars 
(at V$\sim\,$26--27), 
and a continuous distribution of objects at brighter
magnitudes (V$\sim\,$23), with a gap in between. These features are not
reproduced by the simulations, unless a very specific SF history 
(hereinafter the burst scenario) is adopted, as we discuss below. As 
already mentioned, these stars have
been carefully checked and confirmed to be most likely actual single
objects (not extended ones or spurious detections), members of
IZw18.  We have thus considered two different episodes, one of
which could efficiently populate the evolved portions of the tracks
with masses between 12 and 15 \msun (see Fig.~\ref{tracksPC}). This 
corresponds
to force most of the stars in this mass range to be in a post-MS
phase, and is equivalent to consider a burst occurred between 20 and
15 Myr ago. An older longer episode of SF populates the fainter red
giant region.  As an example, in the right hand panels of
Fig.~\ref{sim1fin} we show the CMDs and
LFs obtained assuming a first episode of SF from 1 Gyr to 30 Myr 
ago and a second one, ten times
stronger, from 20 to 15 Myr ago, both with $\alpha$=1.5.  
It can be noticed that in this case all the observational features are 
reproduced pretty well, with deviations of models from the data 
always less than 1\,$\sigma$. 
All the models in better agreement with the data assume quite flat IMFs. 
However, while the second burst can reproduce the observed features only 
if $\alpha\simeq\,$1.5, acceptable distributions are also obtained when 
a Salpeter IMF is adopted for the first episode ($\alpha\simeq\,$2.35), 
with a proper tuning of the SF parameters.  

\subsection{Main body: Simulations with Geneva tracks.}

The set of Geneva tracks with Z=0.0004 (kindly provided by D. Schaerer)
only has models for stars with masses M$\,\ga\,$3 \msun. 
Consequently our simulations cover only the last 0.3 Gyr.
The synthetic CMDs based on this set are all characterized by
a color extension even smaller than that of the Padova models with the same
nominal metallicity. As a general result, these simulations tend to
overpredict the number of bright blue supergiants, while
underpopulating the magnitude range around $V \sim 23.5$. Varying the
IMF and SFR parameters the agreement can be improved, but we did not
find a satisfactory representation of the
data (see A98). An illustrative case is shown in Fig.~\ref{sim3nn}, 
which assumes two episodes of SF: the first
from 300 to 100 Myr 
ago, and a second one started 90 Myr ago and still ongoing.
The IMF slope is $\alpha$ = 2, and the SF rates in the two episodes
are similar to each other. In spite of being one of our best simulations 
with this set of tracks, 
the deficiency of stars with V$\,\simeq\,$23.5 can still be noticed in the LF,
where the maximum deviation is almost 4\,$\sigma$.

Different from the others, the Geneva set with Z=0.001 does span a color range
as large as the one observed in the main body of IZw18. 
In spite of the relatively large value of the $Z$ parameter, the
synthetic blue plume overlaps the observed one when the canonical
E(B--V)=0.04 is adopted. 

As already found with the other sets of stellar tracks, models with only 
one and recent episode of SF activity are definitely inconsistent with the
observed CMD and LF. This can be easily understood from Fig.~\ref{tracksPC1}
where we show the superposition of the observed V--I CMD with the Geneva
Z=0.001 tracks. In order to populate the region at V--I$\,\ga\,$1, 
and fainter than V$\,\simeq\,$25.5, stars of 3--5 \msun ~must have had the
time to evolve off the MS, indicating SF activity earlier 
than $\sim$ 100 Myr ago. Moreover, in the faintest portion of the blue plume 
we find objects of $\simeq$\,5--7 \msun, with ages up to 100 Myr.
For these reasons, simulations with SF starting later than $\sim$ 0.1 Gyr
are inconsistent with the observations, as already shown in the
previous subsection for the Padova set.

Figure~\ref{sim2fin} shows three of the best cases obtained with this set of
tracks under different assumptions for the SF history. In all the cases 
the IMF slope is 1.5. In the left hand panels we plot the result of 
assuming a constant SF over the last 1 Gyr, but stopped 
5 Myr ago. Despite the flat IMF, the synthetic LF is underpopulated around
V$\,\simeq\,$24, with deviations at a 2\,$\sigma$ level,
an inconsistency which worsens with steeper IMFs. Had the SF 
continued in more recent epochs too many bright blue stars would have 
appeared. 
The central panels show a simulation with
two SF episodes, the oldest one started 500 Myr ago and 
stopped 100 Myr ago, while the recent one started 100 Myr ago and still 
active. 
The rate of SF in the first episode is slightly higher 
than in the second one.
In the V vs B--V diagram we can see only the stars born during the most
recent activity. Both the B--V and the V--I LFs deviate from the empirical 
one by at most $\sim$\,3\,$\sigma$, but the color distribution, 
especially for the brightest stars, is not satisfactorily reproduced. 

As already discussed, this particular feature is difficult to
reproduce due to the short lifetimes of massive stars in the post-MS 
phases. The only way to overcome this problem is to 
force the models to populate the bright part of the diagram only with
evolved stars (i.e. with no contribution from the upper MS).
This can be achieved assuming that the more recent SF episode 
stopped fairly long ago (15 Myr ago) so that no stars brighter than 
V$\,\simeq\,$25 can be on the MS. The right hand panels of Fig.~\ref{sim2fin} 
show one of the best cases of this 
type, with the first SF episode from 200 to 30 Myr
ago and the second one from 20 to 15 Myr ago. To obtain enough stars in the
brighter portion of the CMD, we find that the SF rate in the second
burst has been almost 7 times higher than in the old one.
These models reproduce fairly well both the observed distribution of cool and
warm supergiants and the curvature of the upper blue plume in the V--I
CMD. They also reproduce quite well the
observational B--V CMD.

\subsection{Summary of the results for the main body}

The results obtained with the three sets of stellar models 
are consistent with each other in suggesting the overall scenario
for the recent evolution in IZw18. The different values obtained for the 
various parameters depending on the adopted tracks give an estimate of the
theoretical uncertainty still associated with stellar evolution models.
Their relatively small differences support the reliability of our
conclusions. Some of the results are completely independent of the adopted
stellar models, like the flatness of the IMF and the presence of stars with
intermediate ages.

In all the simulations we have found indications of an IMF significantly 
flatter than Salpeter's ($\alpha\,$=2.35). The 
exponents which have turned out to be mostly consistent with the observations 
are in the range $\alpha\,$=1.5--2.0, with some preference for the
flatter extreme of this range. Steeper IMFs look inappropriate also 
in the case of currently ongoing SF,
because they imply too few massive MS stars and too many 
intermediate mass stars, with a consequent overpopulation of the 
faint blue plume. Notice, however, that the derived slope obviously 
refers to the visible range of masses.
At the distance of IZw18, nothing can be inferred with our method on the 
IMF of stars less massive than $\sim$ 2 \msun.

Given the relatively short lookback time of the empirical CMDs of 
IZw18,
we have considered a star formation activity distributed over one or at most 
two episodes with a regime constant or exponentially decreasing with time.
We should recall that two is possibly the maximum number
of SF bursts allowed by the extremely low metallicity of this galaxy (e.g.
Kunth, Matteucci, \& Marconi 1995).
In no case have we been able to reproduce the observed CMDs and LFs with one 
single episode of SF started 
more recently than 0.1 Gyr ago. This rules out, beyond any reasonable doubt, 
that IZw18 has started very recently to form its first stars.

A single SF episode can reproduce rather well the data if extending over
a sufficiently long period of time ($\ga$ 0.2 Gyr). With an IMF slope of 1.5
we derive typical SF rates of $\sim 6\times10^{-3}$ \msuny ~for stars more
massive than 1.8 \msun. If the SF episodes sampled by the resolved stars 
in IZw18 are two, we find a better agreement between synthetic predictions 
and empirical data especially when the younger episode is relatively old and 
7--10 times stronger than the previous one. To reproduce the observed features,
the younger episode must have occurred between $\sim$ 20 and 15 Myr, with a 
SFR of $\sim 3 (6) \times10^{-2}$ \msuny ~for the Geneva (Padova) tracks.
The older episode can have started any time between 1 and 0.2 Gyr ago, and 
continued until approximately 30 Myr ago. An earlier stop of the latter
SF activity would lead to an underpopulated blue plume at the faint 
magnitudes. If we want instead the SF in IZw18 to have taken place until 
recently, not only in the densest unresolved regions (Kunth et al. 1995; De 
Mello et al. 1998; Izotov \& Thuan 1998; Van Zee et al. 1998), but also in 
the resolved field, the best agreement is attained if the most recent of 
these two episodes has occurred from 0.1 Gyr ago at a rate of 
2--5 $\times10^{-3}$ \msuny. In this case, however, as well as in the 
single-episode case, the yellow/red supergiants observed in IZw18 are not reproduced
by the models. For this reason we definitely prefer the burst scenario with
the intense SF episode between 20 and 15 Myr ago.

To evaluate the actual SFR in IZw18, the value obtained 
from the synthetic CMDs must be extrapolated from the lower mass limit 
adopted in the simulation (m$_{\rm low}$ = 1.8, 2 \msun ~for the Padova 
and Geneva sets respectively) to the physical lower mass cutoff. Since
the IMF at the low mass end is still highly uncertain (Larson 1998; 
Leitherer 1998) both in the slope and in the lower mass cutoff, the 
extrapolations have been performed exploring a few simple cases. For the 
lower mass cutoff, we have adopted the value of 0.1 \msun.
If $\alpha$=1.5 over the whole mass range, the extrapolation leaves basically
unaltered the SFRs quoted above. Alternatively, if a Salpeter slope is adopted 
below m$_{\rm low}$ the corrected SFR amounts to 1.4 times the values quoted 
above.

Since the size of IZw18 is estimated to be 840 $\times$ 610 pc$^2$ (DH90), 
the rates presented above and corrected for the IMF extrapolation 
become on average 1--2 $\times$ 10$^{-2}$
M$_{\odot}$yr$^{-1}$kpc$^{-2}$ in the cases of one or two SF episodes. Only
when the second episode stops as early as 15 Myr ago, its SFR can be as
high as 6--16 $\times$ 10$^{-2}$ M$_{\odot}$yr$^{-1}$kpc$^{-2}$,
depending on the adopted IMF and evolutionary tracks.

\subsection{Simulations for the secondary body}

The fiducial stars populating the CMD of IZw18 companion are so few 
(58), that the comparison with the corresponding synthetic diagrams is
inevitably affected by small number statistics. Besides, only the V, B--V
CMD is available for this object, thus sensibly reducing the lookback time, 
and in general the available
constraints to discriminate between different evolutionary scenarios.
Nevertheless some interesting conclusions can still be drawn, thanks 
to the circumstance that all the sets of stellar tracks favor the same 
overall scenarios for its star formation history. For this reason, in
the following we show only the results for the Padova set of tracks.

As illustrated in Section 3.2, the blue plume of the secondary body is
1.5--2 mags fainter than that of the main body. This is not necessarily
a signature of an older stellar population, since we have seen in the
previous sections that the brightest blue stars in the main body are mostly 
post-MS objects, much brighter than their MS progenitors.
As visible in Fig.~\ref{tracksWF3}, the red portion of the blue plume
fainter than V=25.5 can be populated either by stars of approximately
4--6 \msun in the blue loop, or by more massive stars still on the MS.
As a consequence, the observed magnitude distribution of the 
stars in the companion can be reproduced either with a quite young SF episode
(started around 50 or less Myr ago) or with a rather old one (started 
around 200 Myr ago). Nonetheless the color of the blue plume is (slightly) 
better reproduced by the older scenario.

Figure~\ref{simsec1fin} shows the synthetic diagrams obtained for a SF started 
10 Myr ago in the top and third panel, and a SF started 150 Myr
ago in the second panel. Their luminosity functions are shown in the bottom
panel (solid, dotted and dashed lines, for the top, second and third CMD,
respectively). In order to compensate for the higher number 
of massive young stars in the top case, its IMF slope is steeper than in the
second case, 2.6 and 1.5 respectively. It can be seen that both models give
a fair representation of the data.
The third panel (and dashed LF) shows what happens to the top panel model 
if one only changes the adopted IMF slope from 2.6 to 1.8. 
Too many massive blue supergiants populate the top of the blue plume, 
making it far too bright, and, correspondingly, too few MS stars populate 
its faint end.
Similar results are obtained with the other sets of tracks, though with
somewhat different values for the parameters, reflecting the 
different lifetimes in the various evolutionary stages. For example, slightly
earlier starts for the SF activity and steeper IMF slopes are derived with
the Z=0.001 Geneva sequences. 

In the case of the secondary 
body where the observational constraints are modest, the range of acceptable 
values for the parameters is larger than for the main body. In addition,
as shown in Fig.~\ref{simsec1fin}, it is difficult to 
disentangle the contribution of the IMF and of the SFR to the observed 
stellar distribution. 
From the hundreds simulations performed on the secondary body, we 
believe that no quantitative information can be derived on its IMF slope.
The range of acceptable slopes for the IMF is large (1.5--3.0), but 
the slopes leading to diagrams in better agreement with the data are peaked 
at $\alpha$=2.2, somewhat flatter than Salpeter's ($\alpha$=2.35), and 
definitely steeper than the slopes required for the main body. Besides, the 
trend that the more recent the start of the SF episode, the steeper the IMF 
is confirmed by all models. With $\alpha$ flatter than 2, a SF started more
recently than $\sim$30 Myr ago and still active has to be excluded, and one
started as early as 0.15--0.20 Gyr is preferable. For steeper IMFs,
SF activities started as late as 10 Myr ago and still ongoing can be
appropriate to interpret the observed features.

% Another result, common to all the three sets of stellar models, is that
% the SF cannot have stopped too early. The second panel of
% Fig.~\ref{simsec2fin} shows a simulation in which the SF activity 
% started 50 Myr ago and stopped
% 10 Myr ago. As a result, the turn-off is fainter than V = 26, and a clear
% gap appears in the blue plume at the separation between MS and post-MS stars.
% This case is based on the Geneva tracks with Z=0.001 and assumes
% $\alpha$=2.6, but very similar results
% are obtained also with the other tracks and with different IMF and SF
% regimes whenever the activity is forced to stop as early as 10 Myr ago.
% These results indicate that the stars in the secondary body are
% possibly younger on average than those in the main body, despite the lower 
% brightness of its blue plume.

The rate of SF is obviously inversely proportional to the duration of the
activity (since the number of generated stars still visible is given by
the data). Considering the extrapolation from m$_{low}$ to the lower physical 
mass cutoff 0.1 M$_{\odot}$ either with a single-slope IMF with $\alpha$=2.2 
or with Salpeter's slope, the derived SFRs must be corrected 
by a factor of 2.5 or 2.9, respectively. Thus for young SF episodes, 
occurring in the last 10--50 Myr, the average rate is 
2--5 $\times 10^{-3}$M$_{\odot}$yr$^{-1}$,
depending on the adopted stellar tracks.
For a SF activity started as early as 0.2 Gyr ago, the average rate 
is lower, 1--2 $\times$ 10$^{-3}$ M$_{\odot}$yr$^{-1}$. 
In terms of rate per unit area, these values translate into 0.7--1.7 
$\times$ 10$^{-2}$ and 3.4--6.7 $\times$ 10$^{-3}$ 
M$_{\odot}$yr$^{-1}$kpc$^{-2}$, 
respectively, once a size of 850 $\times$ 350 pc$^2$ is assumed for the 
secondary body (DH90). Therefore, the average SF rate in the secondary body 
has been similar or $\sim$3 times lower than in the main body,
depending on the preferred scenario.

\section{Discussion and Conclusions}

We have studied the SF history in IZw18, with the main goal of trying to
disentangle the long standing question of whether or not this system is 
experiencing now its first burst of star formation. Other investigators have 
already examined this question and provided contradicting answers.
To mention just one recent example: Kunth et al. (1995) inferred from a 
series of chemical evolution models that
IZw18 can have experienced at most two SF bursts, each of which with duration
no longer than 10--20 Myr, whereas Legrand \& Kunth (1998, hereinafter LK) 
argue from a
spectro-photometric-chemical model that the observed metal abundances and
colors can be better explained in terms of a very low SFR
(10$^{-4}$M$_{\odot}$yr$^{-1}$), continuous during 16 Gyr, with
burst occurrence (and SFR 100 times stronger) only in the last 50 Myr.

Other authors have argued in favour of a relatively recent onset of
the SF activity in IZw18. Both HT95 and D96 suggest a continuous, and still
ongoing SF over the last 30--50 Myr, as deduced comparing the CMDs of
the resolved stellar population on WFPC2 images with isochrones.
From the kinematic analysis of ionized gas, Martin (1996) found a
bipolar bubble with a lobe more evident in the SE than in the NW part of the
galaxy. Its dynamical evolution and photometric properties are
well described by a continuous SF episode started 15--30 Myr ago at a
rate of $\sim$ 0.02 \msuny. On the other hand, in the literature there are
some clues of an older SF activity in IZw18; for instance, from the 
comparison of the C/O ratio with
predictions of chemical evolution models, Garnett et al. (1997) suggest
a SF episode as long as a few hundreds Myr.

Our approach is to infer the SF history of the galaxy from the CMDs and
LFs of its stars resolved by HST photometry. As already mentioned in the
previous sections, this method does not examine the denser, unresolvable
regions where some SF has certainly occurred at very recent epochs as
demonstrated by the presence of several HII regions.
The derived V, V--I and V, B--V
diagrams have been interpreted in terms of SF and IMF by means of
theoretical simulations.
In comparison with other galaxies examined with the same method, it is
more difficult to derive strict constraints on the SF history of the IZw18 
system, because its larger distance makes much smaller the number of 
resolved stars and consequently poorer the statistical significance of 
the results, especially for the secondary body.
Nonetheless, in spite of this problem and of the difficulties described
above to fully reproduce all the observed features of the galaxy, 
the comparison of all the synthetic CMDs and LFs with the corresponding
data, has led to quite firm indications on the overall
properties of the evolution of IZw18.

It is clear from the results presented in the previous 
sections that in no way can a single SF episode started only a few tens of Myr 
ago reproduce the observed features of the faint blue plume of the main body. 
The SF in IZw18 must have been already active at least 100 Myr (but
more likely 500 Myr) ago to provide 
all the observed faint stars, both blue and red. This same conclusion is 
reached with all the available sets of stellar evolution tracks and is 
therefore independent of the adopted models; it can then be considered quite 
firm. The overall scenario for the SF history of IZw18 is thus an almost 
constant SF activity from 1 Gyr up to $\sim$30 Myr ago coupled with a burst 
almost ten times stronger around 15--20 Myr ago:
the oldest stellar population is practically concentrated in the SE part
of the galaxy, while the other stars are
both in the NW and SE inner dense regions (see A98).

The presence of relatively old stars excludes one of the two alternative 
scenarios proposed by Kunth et al. (1995), which allows for only 
one ongoing episode started a few Myr ago. The alternative case, of two
separate episodes is instead compatible with our results. At first glance, 
our results seem also in agreement with LK's scenario of an almost continuous 
star formation activity. However, the average SF 
during the epochs covered by our analysis has turned out to be 
$\sim 10^{-2}$M$_{\odot}$yr$^{-1}$, two orders of magnitude
higher than the {\it low} level predicted by LK's model. Our rate is instead
close to what LK attribute to the current burst. On the other
hand, the duration of the SF activity is much longer in our scenario than in 
LK's burst. We do find
that a burst is likely to have occurred at roughly LK's burst rate, but
in a shorter time interval (from 20 to 15 Myr ago in our scenario, from
50 Myr ago until now in LK's). Thus, our quantitative
conclusions do not necessarily agree with LK's values. This of course
does not exclude that a continuous SF activity has taken 
place throughout the galaxy lifetime, but it should have had an intensity 
quite lower than in their model, to compensate the longer duration of
the recent interval at high rate.
Our derived SF history is instead in agreement with Martin (1996) and Garnett 
et al. (1997) results. In particular, both the epoch and the level of the SFR 
in the most recent episode of our burst scenario agree with Martin's (1996)
finding. Thus, from the study of the resolved stars in IZw18 we find support 
to the idea that this episode of SF powered the bipolar bubble and possibly 
a galactic outflow.

For a direct comparison of the derived SFRs in IZw18 with those of other 
dwarfs, it is more physically meaningful to consider the rate per unit area.
In these units the main body of IZw18 has a SFR 10$^{-2}$--10$^{-1}$ 
M$_{\odot}$ yr$^{-1}$ kpc$^{-2}$, and the secondary body a SFR
3--10 $\times 10^{-3}$ M$_{\odot}$ yr$^{-1}$ kpc$^{-2}$.
The SFRs derived with the same method for Irregular Galaxies of the Local
Group (e.g. Tosi 1998) are in the range 10$^{-4}$--10$^{-2}$ 
M$_{\odot}$ yr$^{-1}$ kpc$^{-2}$,  while for the extremely active dIrr NGC~1569
(G98) the estimated recent SFR is between 4 and 20 M$_{\odot}$ yr$^{-1}$ 
kpc$^{-2}$ depending on the adopted IMF (2.35$\leq\alpha\leq$3.0).
In the solar neighborhood the present 
SFR is in the range (0.2--1)$\times$10$^{-2}$ M$_{\odot}$ yr$^{-1}$ kpc$^{-2}$ 
(Tinsley 1980b; Timmes, Woosley, \& Weaver 1995).
We can thus conclude that IZw18 shows a mean SF activity comparable to that 
of the region around the sun and that of the most active Local Group 
Irregulars. As a consequence, its SFR falls short of $\sim$ 2 orders of 
magnitude to make IZw18 a local counterpart of the faint blue galaxies,
according to Babul \& Ferguson (1996) model.

In all the approximately 500 simulations performed for the CMDs of IZw18 we 
have found indications of an IMF significantly flatter than Salpeter's
($\alpha$=2.35). The 
exponents which have turned out to be mostly consistent with the observations 
are in the range $\alpha\,$=1.5--2.0, with some preference for the
flatter extreme of this range. 
This is the first galaxy in our sample showing such a significant evidence
in favour of a flat IMF. All the others analyzed by us with the same approach 
(DDO~210, NGC~1569, NGC~3109, NGC~6822, Sex~B, WLM) turned out to have IMF 
slopes 
close to Salpeter's or slightly steeper, in agreement with the current general
belief (e.g. Leitherer 1998) of a roughly universal IMF in irregular galaxies.
Besides, the global LF of IZw18 seems steeper and not flatter than those of
other irregulars (see Sect.3.1). For this reason we have examined with 
particular attention all the alternatives, to evaluate the 
possibility that a more standard IMF could be acceptable with other parameter 
combinations. However, in no case have we been able to reproduce the observed 
CMDs and LFs if all the stars were born following Salpeter's IMF.
As mentioned in Sect. 4.1, we obtained acceptable results adopting a Salpeter 
slope in the older SF episode, 
but a flat IMF in the most recent generation seems
required by the data. Could this peculiarity be due to the extremely low 
metallicity of IZw18, following the old suggestion (e.g. Terlevich 1985; 
Melnick \& Terlevich 1987) that the lower the metallicity, the flatter the 
IMF of massive stars ? This interesting possibility is however contradicted 
by the possible trend of the older metal-poorer episode more consistent 
with a steeper IMF than the younger metal-richer one 
and by the circumstance that the 
secondary body seems to have a more standard IMF (with $\alpha \sim 2.2$), 
despite the same low metallicity of IZw18. 

It is worth to stress here also the possible correlation existing between the
SF activity in IZw18 and the stellar production in its companion system.
Many papers in the literature consider the stellar population in component C
older than that in the main body. Different studies on its resolved stellar
population (D96), ionized gas (Izotov \& Thuan 1998), integrated
colors and nebular spectrum (Van Zee et al. 1998) indicate ages spanning
from 100 to 300 Myr, all consistent with our older scenario for this
minor system. The interpretation of the empirical CMD and LF in the 
secondary body is however 
much less constrained than for the main body, due to the small number of 
observed stars. Indeed we find consistency with the data also adopting 
a recent and still ongoing SF activity, provided that the IMF exponent is
steep enough.
Deeper and more accurate data would be necessary to derive
tighter conclusions.

On somewhat speculative grounds, the general trend may be that 
of a SF propagating from the secondary body 
through the NW part of IZw18 to its SE component, where some HII regions and 
the brightest young clusters are concentrated and still visible 
(HT95; D96; Izotov \& Thuan 1998). Admittedly, the spatial correlation between 
stellar production in different regions of IZw18 and in its companion 
system is not strong. However, in the case of recent onset of the SF activity 
in the minor system (50--10 Myr ago), its starting epoch is similar to that
(20--15 Myr ago) of the stronger burst in the main body.
Also the SFRs are roughly comparable: 
$\sim$0.1 M$_{\odot}$yr$^{-1}$kpc$^{-2}$ for the intense burst in IZw18, and 
$\sim$0.02 M$_{\odot}$yr$^{-1}$kpc$^{-2}$ for the companion system. Besides,
the stars 
in the main body generated during the strong burst (red and yellow supergiants) 
are possibly located preferentially in the NW part, near the companion: this 
burst in IZw18 might thus have been triggered by gravitational interactions.

To conclude, from the analysis of the resolved stellar population in IZw18 
our major results are the following: the age of the older stars seen in the 
main body reaches from a few hundreds Myrs up to $\sim$1 Gyr; therefore a 
single recent 
episode of SF is ruled out. Our preferred scenario for the SF history 
is the burst scenario, consisting of two episodes, the younger one having 
occurred between 15 and 20 Myr ago at a rate 7-10 times higher than in the 
previous activity. This refers to our analyzed field, and not to the denser 
regions where an ongoing SF 
activity shows up through the HII regions and unresolved star clusters.
The SFRs that we derive for the main body of IZw18 are similar to
those of nearby irregulars and the solar neighborhood. The IMF,
instead, appears to be significantly flatter than in any of these normal 
galaxies. This is especially true for the second of the two episodes, the real 
burst. The IMF in the secondary body appears instead to be less extreme, with
a likely slope of $\alpha\simeq\,$2.2.
  
{\bf Acknowledgements}
We warmly thank Mark Clampin, Antonella Nota and Marco Sirianni who have been 
of invaluable help. We are deeply indebted with Daniel Schaerer for having 
computed for us the stellar models with Z=0.0004, to Peter Stetson for 
providing some of his software and to Manuela Zoccali for help in using it. 
Paolo Montegriffo has also helped a lot with his unusual skill for photometric 
reduction in very crowded fields. Useful conversations with Claus Leitherer, 
Daniel Schaerer and Michele Bellazzini are also acknowledged. 
We are grateful to the anonymous referee for his/her useful comments 
and suggestions which contributed to improve the paper. Part of this 
work has been funded by the Italian Space Agency ASI.

\clearpage

\begin{deluxetable}{cccccl}
\footnotesize
\tablecaption{HST/WFPC2 Archival Data of IZw18}
\tablewidth{0pt}
\tablehead{
\colhead{Filter}  &
\colhead{Camera}  &
\colhead{PI}  &
\colhead{Epoch}  &
\colhead{$ \stackrel{\textstyle {\rm Exposure}}{ ({\rm sec}) }  $} &
\colhead{Root Name}
}

\startdata

F336W & PC & Hunter & 29 Oct 1994 & 1400 & u2cg0101t,u2cg0102t,u2cg0103t 
\nl
F439W & PC & Dufour & 01 Mar 1995 & 1000 & u2f90303t$^*$,u2f90304t$^*$ \nl
F450W & WF3 & Dufour & 03 Nov 1994 & 2300 & u2f90102t$^*$,u2f90103t$^*$ \nl
F469N & PC & Hunter & 31 Oct 1994 & 2300 & u2cg0401t,u2cg0402t,u2cg0403t 
\nl
F502N & WF3 & Dufour & 02 Nov 1994 & 2300 & u2f90203t$^*$,u2f90204t$^*$ \nl
F555W & PC & Hunter & 31 Oct 1994 & 2200 & u2cg0201t$^*$,u2cg0202t$^*$,
u2cg0203t$^*$ \nl
      & PC & Dufour & 01 Mar 1995 & 600 & u2f90301t$^*$,u2f90302t$^*$ \nl
      & WF3 & Dufour & 03 Nov 1994 & 2300 & u2f90104t$^*$,u2f90105t$^*$ \nl
F656N & PC & Hunter & 03 Nov 1994 & 1400 & u2cg0501t$^*$,u2cg0502t$^*$,
u2cg0503t$^*$ \nl
F658N & WF3 & Dufour & 03 Nov 1994 & 2300 & u2f90205t$^*$,u2f90206t$^*$ 
\nl
F675W & PC & Dufour & 01 Mar 1995 & 1000 & u2f90305t,u2f90306t \nl
F702W & WF3 & Dufour & 02 Nov 1994 & 1800 & u2f90101t,u2f90201t,u2f90202t 
\nl
F814W & PC & Hunter & 30 Oct 1994 & 2200 & u2cg0301t$^*$,u2cg0302t$^*$,
u2cg0303t$^*$ \nl

\enddata
\end{deluxetable}

\clearpage

\begin{deluxetable}{ccccccccc}
\footnotesize
\tablecolumns{9} 
\tablewidth{0pt}
\tablecaption{Photometric completeness as a function of magnitude}
\tablehead{
\colhead{} &
\multicolumn{2}{c}{IZw18: PC Images} &
\colhead{} &
\multicolumn{2}{c}{IZw18: PC Images} &
\colhead{} &
\multicolumn{2}{c}{Comp.: WF3 Images}  \\
\cline{2-3} \cline{5-6} \cline{8-9} \\
\colhead{Magnitude}  &
\colhead{$ \stackrel{\textstyle {\rm Deep}}{{\rm F555W}} $}  &
\colhead{$ \stackrel{\textstyle {\rm Deep}}{{\rm F814W}} $} &
\colhead{} &
\colhead{$ \stackrel{\textstyle {\rm Shallow}}{{\rm F555W}} $}  &
\colhead{$ \stackrel{\textstyle {\rm Shallow}}{{\rm F439W}} $}  &
\colhead{} &
\colhead{$ \stackrel{\textstyle {\rm Deep}}{{\rm F555W}} $}  &
\colhead{$ \stackrel{\textstyle {\rm Deep}}{{\rm F450W}} $}  
}

\startdata

   $<$21.0     &     100    &     100    & &     100    &     100    & 
&     100    &    100    \nl
21.0$\div$22.0 &     100    &     100    & &  90$\pm$31 &  95$\pm$22 & 
&     100    &    100    \nl
22.0$\div$23.0 &  89$\pm$16 &  99$\pm$4  & &  83$\pm$22 &  93$\pm$16 & 
&     100    &    100    \nl
23.0$\div$24.0 &  85$\pm$12 &  90$\pm$10 & &  79$\pm$18 &  83$\pm$16 & 
&     100    &    100    \nl
24.0$\div$25.0 &  73$\pm$13 &  74$\pm$12 & &  65$\pm$23 &  64$\pm$23 & 
&  83$\pm$26 & 93$\pm$18 \nl
25.0$\div$26.0 &  52$\pm$13 &  47$\pm$12 & &  36$\pm$20 &  20$\pm$18 & 
&  80$\pm$29 & 84$\pm$27 \nl
26.0$\div$27.0 &  20$\pm$8  &  13$\pm$9  & &   5$\pm$6  &     0      & 
&  68$\pm$29 & 37$\pm$35 \nl
27.0$\div$28.0 &   5$\pm$7  &      0     & &     0      &    ...     & 
&     0      &  4$\pm$13  \nl
28.0$\div$29.0 &       0    &     ...    & &    ...     &    ...     & 
&    ...     &     0      \nl

\enddata
\end{deluxetable}

\clearpage

\figcaption{Deep combined image of IZw18 on a logarithmic scale in the WFPC2 
F555W filter. The target is centered on the PC camera and the displayed field 
of view is 31\farcs5$\times\,$31\farcs5. The pixel 
scale corresponds to 0\farcs045 pixel$^{-1}$ and the total 
integration time is 6,600 s. 
\label{VimagePC}}

\figcaption{IZw18 companion system displayed in logarithmic scale on the 
combined F555W image of the WF3 camera for a total integration time of 4,600 
s. The field of view considered is 
28\farcs6$\times\,$20\farcs1 and the resolution 
0\farcs1 pixel$^{-1}$. The two objects inside a black circle
on the right are the two bright red stars survived in the final observed 
CMD: because of their brightness together with their distance from the 
central stellar condensation, it could be possible that they do not belong 
to the system.
\label{VimageWF3}}

\figcaption{Aperture correction from the deepest F555W and F814W PC images 
as a function of the aperture radius. See text for symbols and details.
\label{apcorr}}

\figcaption{Photometric errors versus calibrated magnitude as given by 
DAOPHOT for IZw18 photometry in the combined PC frames. From top to 
bottom: the deeper F555W and F814W frames from Hunter's data, 
the shallower F555W and F439W images from Dufour's data.
\label{errPC}}

\figcaption{Same as Fig.~\ref{errPC} but for IZw18's companion in the 
combined F555W (top panel) and F450W (bottom panel) images of the WF3 camera. 
\label{errWF3}}

\figcaption{Comparison of our and HT95 photometries in the F555W (top panel) 
and F814W (bottom panel) filters: in these two diagrams we plotted all the 
stars we re-measured from the list of HT95 using our photometric procedures 
and calibrations. The zero-point offsets resulting from least-squares fits
are: $\Delta$V$\sim$\,--\,0.22 and $\Delta$I$\sim$\,--\,0.14.
\label{photcomp}}

\figcaption{CMD in the F555W and F814W filters for IZw18 as observed in the 
PC field of view: a) 444 objects with $\chi^2\!<\,$3 and 
-1$<\!sharpness\!<$1; 
b) subsample after selecting the objects with photometric error 
$\sigma_{DAO}\!<\,$0.2 in both filters and after having cleaned the CMD 
from uncertain detections (247 stars).
\label{VIcmdPC}} 

\figcaption{Same as Fig.~\ref{VIcmdPC} but for the F555W and F439W 
bands of the PC camera (respectively 267 and 106 objects).
\label{BVcmdPC}}

\figcaption{CMDs of IZw18 compared to the Padova tracks with Z=0.0004: panel a) 
refers to the V vs V--I, panel b) to the V vs B--V.  The stellar
mass of each track is given in M$_{\odot}$. The right vertical axis is the 
absolute magnitude in the F555W filter.
\label{tracksPC}}

\figcaption{Luminosity function of stars in IZw18 present in the 
$\sigma_{DAO}\!<\,$0.2 CMD: 
panel a) refers to the V vs V--I diagram, 
thus to the deeper F555W image, while panel b) to the F555W shallower frame 
of the V vs B--V. Error bars correspond to the rms.
\label{PCLFs}}

\figcaption{CMD of IZw18's companion obtained from the F555W and F450W WF3 
frames: a) and b) as in Fig.~\ref{VIcmdPC} (respectively 109 and 58 stars).
\label{BVcmdWF3}}

\figcaption{Luminosity function of stars with $\sigma_{DAO}\!<\,$0.2
in IZw18's companion.
\label{WF3LF}}

\figcaption{Same as Fig.~\ref{tracksPC} but for the V vs B--V of IZw18 
companion galaxy.
\label{tracksWF3}}

\figcaption{Synthetic CMDs and LFs with Padova tracks for the main body of 
IZw18: Z=0.0004.
The left hand panels show the case of one SF episode started 10 Myr ago and
still active at constant rate, with an IMF slope of 3.0.
From top to bottom: V vs V--I synthetic CMD;
its LF (solid line) compared with that derived from the empirical V vs V--I 
CMD (dots); V vs B--V synthetic CMD; its LF (solid line) compared with that 
derived from the empirical V vs B--V CMD (dots).
The central panels refer to the case of one SF episode started 1 Gyr ago,
with exponentially decreasing SFR ($\tau$=0.5 Gyr), still active, and
with $\alpha$ = 1.5.
The right hand panels correspond to two SF episodes, one started 1 Gyr ago
and stopped 30 Myr ago, and the second started 20 Myr ago and lasted only
5 Myr. Also in this case the IMF is $\alpha$=1.5.
\label{sim1fin}}

\figcaption{Synthetic CMDs and LFs with Geneva tracks for the main body of 
IZw18: Z=0.0004. The adopted IMF slope is 2.0, and the SF has occurred in two 
episodes, from 300 to 100 Myr ago and from 90 Myr ago on. 
\label{sim3nn}}

\figcaption{V, V--I CMD of IZw18 compared to the Geneva tracks with Z=0.001.
Symbols and coordinates are as in Fig.\ref{tracksPC}.
\label{tracksPC1}}

\figcaption{Synthetic CMDs and LFs with Geneva tracks for the main body of 
IZw18: Z=0.001.
The left hand panels show the case of a constant SF activity from 1 Gyr to 5
Myr ago. 
The central panels correspond to two SF episodes: from 0.5 to 0.1 Gyr ago,
and from 0.1 Gyr ago until now. 
The right hand panels refer to two SF episodes: from 0.2 Gyr to 30 Myr ago,
and from 20 to 15 Myr ago. 
The IMF slope is $\alpha$=1.5 in all the cases. 
\label{sim2fin}}

\figcaption{Synthetic CMDs and LFs for the secondary body of 
IZw18 based on the Padova tracks with Z=0.0004. The top CMD corresponds to 
a constant SF episode started 10 Myr ago, the central CMD to a relatively old 
exponentially decreasing SF, started 150 Myr ago, and the lower CMD to
the same model as in the top panel, but with flatter IMF.
The adopted IMF slope is indicated. The LFs (solid line for the top CMD, 
dotted line for the central one, and dashed line for the lower CMD) are 
compared with the empirical one in the bottom panel. 
\label{simsec1fin}}

%\end{document}
%%%%%%%%%%%%%%%%%%%%%%%%%%%%%%%%%%%%%%%%%%%%%%%%%%%%%%%%%%%%%%%%%%%%%

\clearpage

\epsscale{0.8}
%\plotone{VimagePC.eps}
\plotone{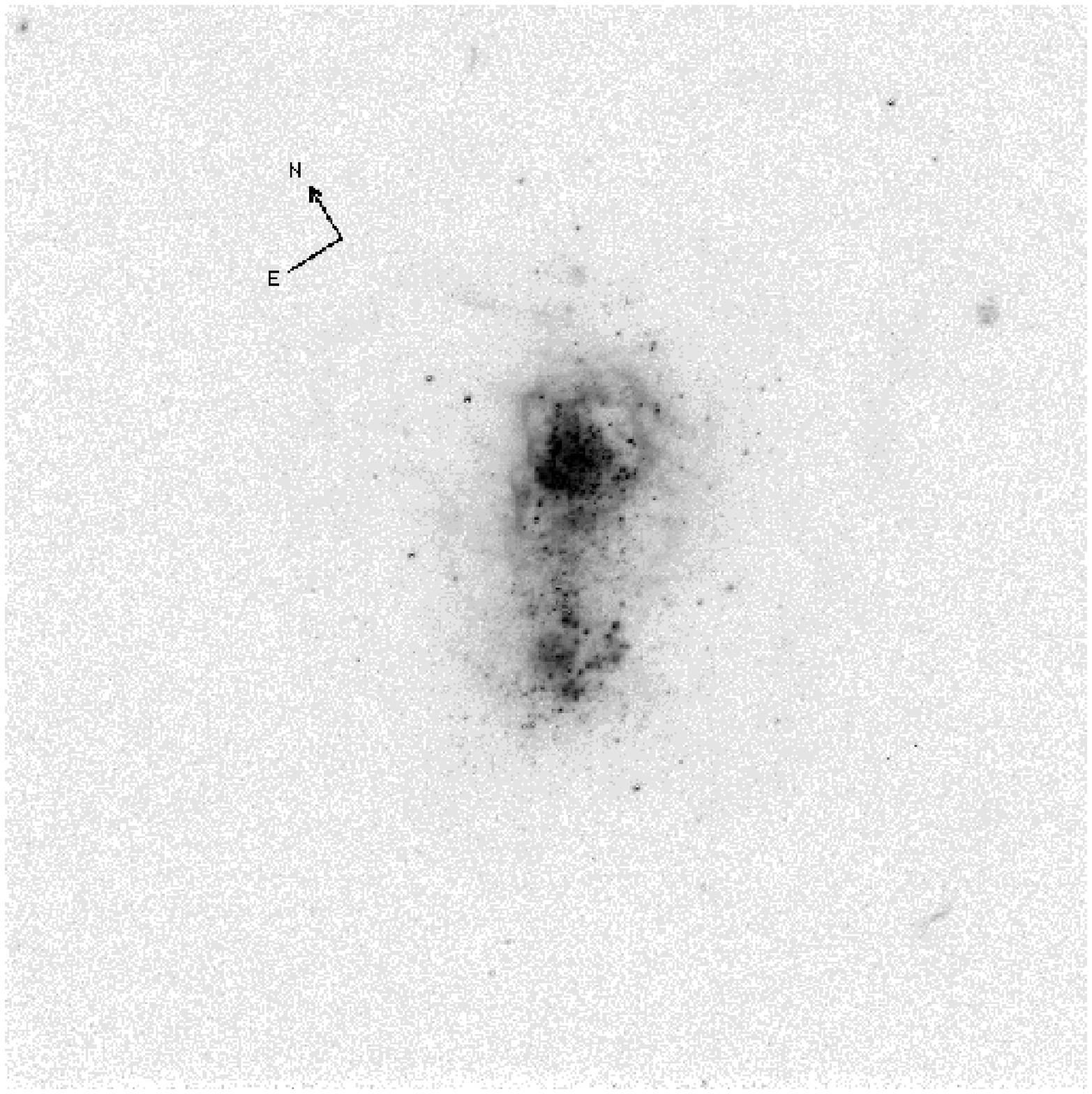}
%1

\clearpage

%\plotone{VimageWF3.eps}  
\plotone{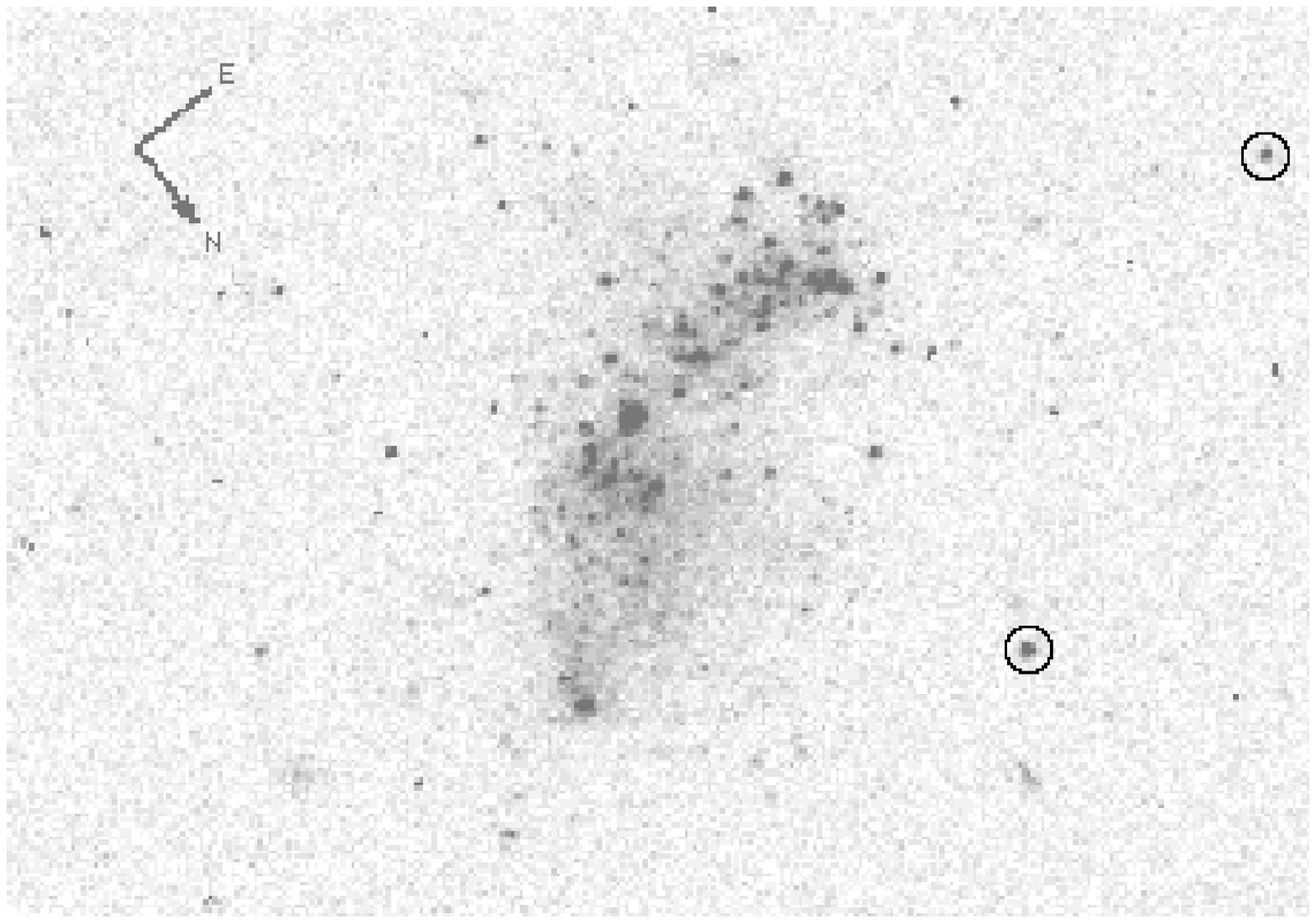}  
%2

\clearpage

\epsscale{0.7}
%\plotone{apcorr.eps}
\plotone{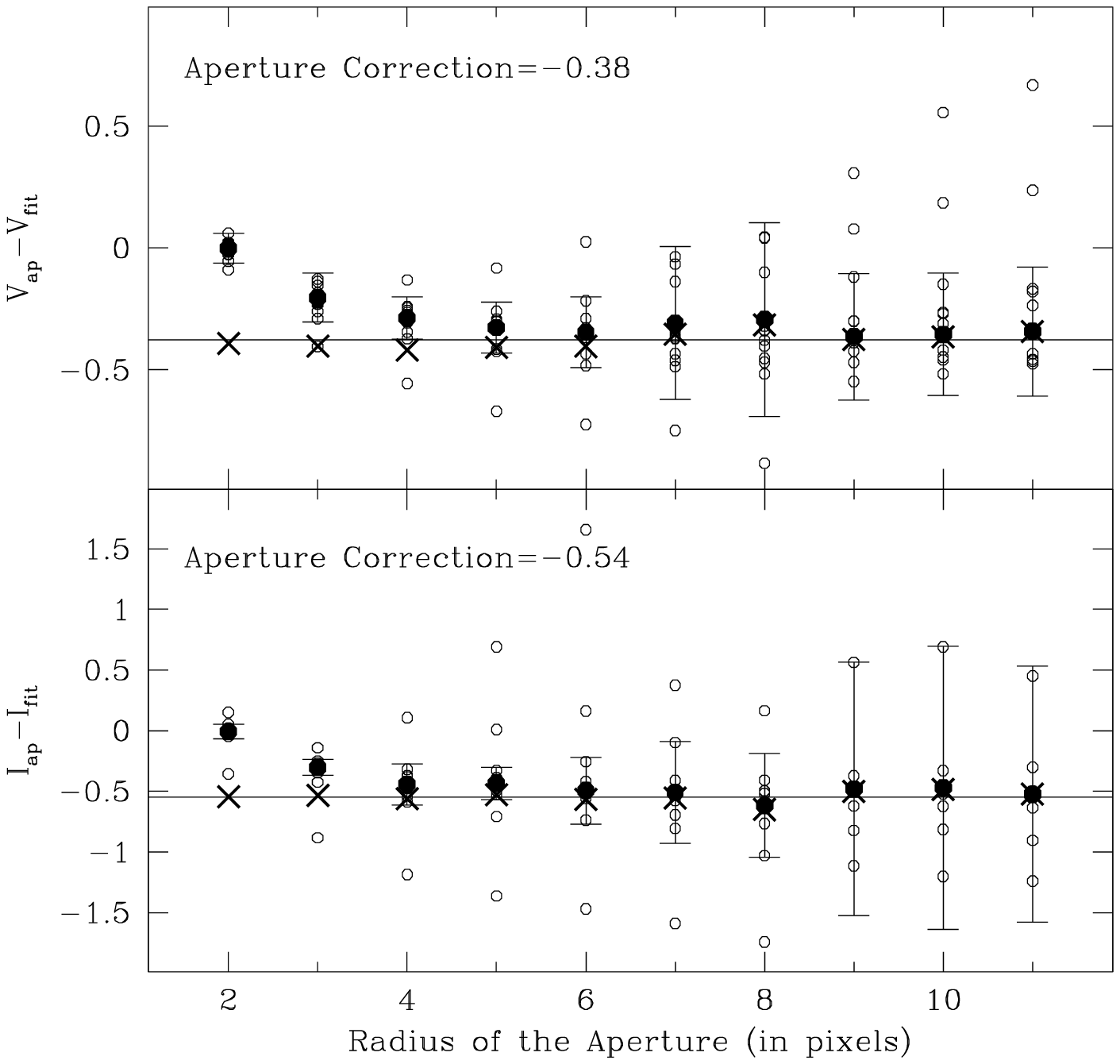}
%3

\clearpage

\epsscale{0.6}
%\plotone{errPC.eps}
\plotone{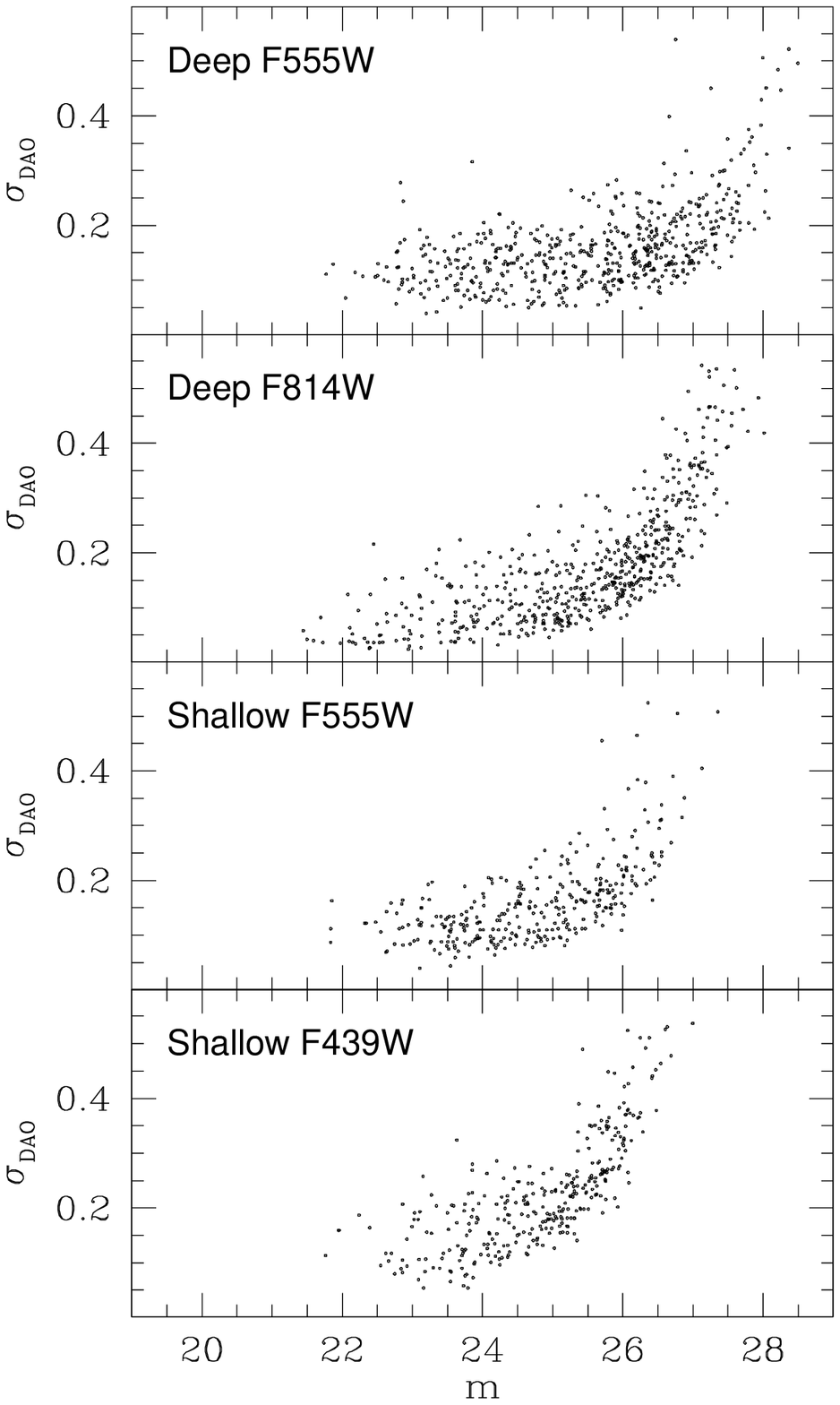}
%4

\clearpage

%\plotone{errWF3.eps}
\plotone{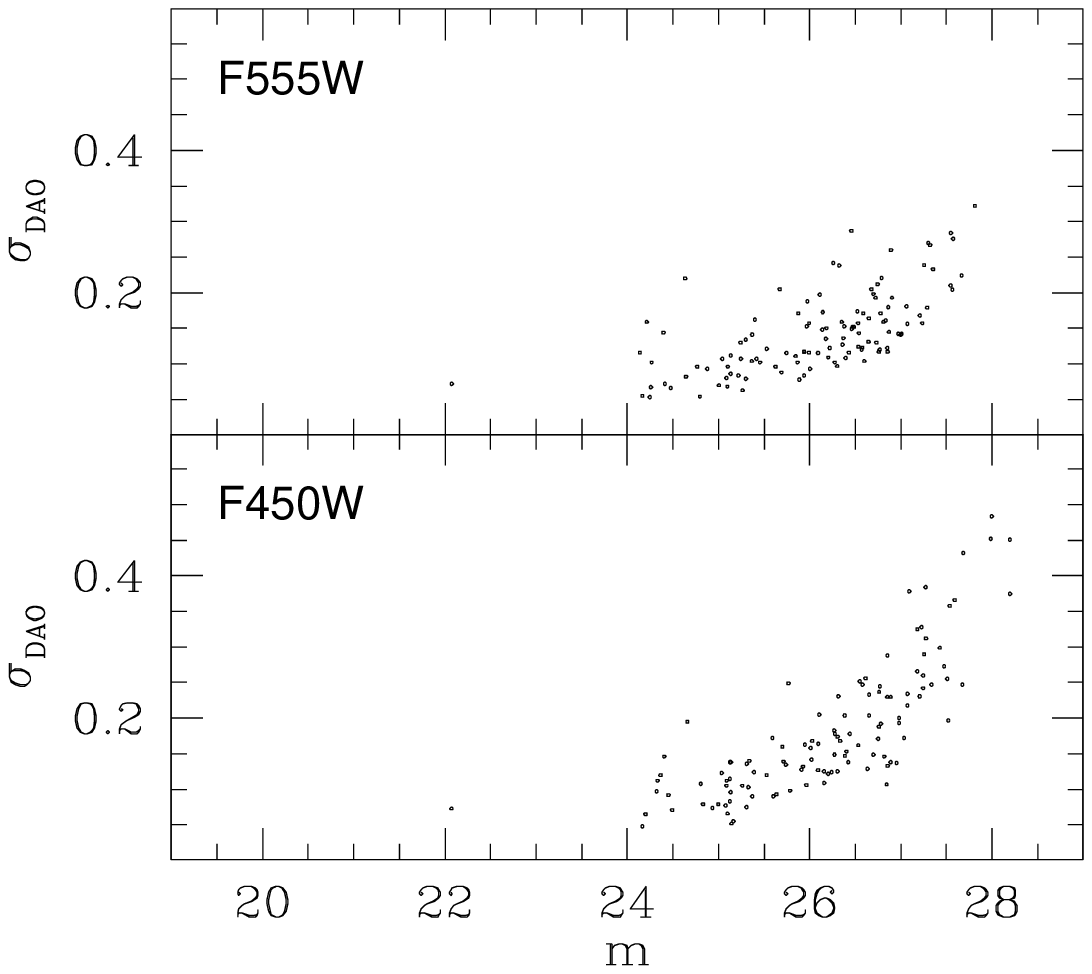}
%5

\clearpage

%\plotone{photcomp.eps}
\plotone{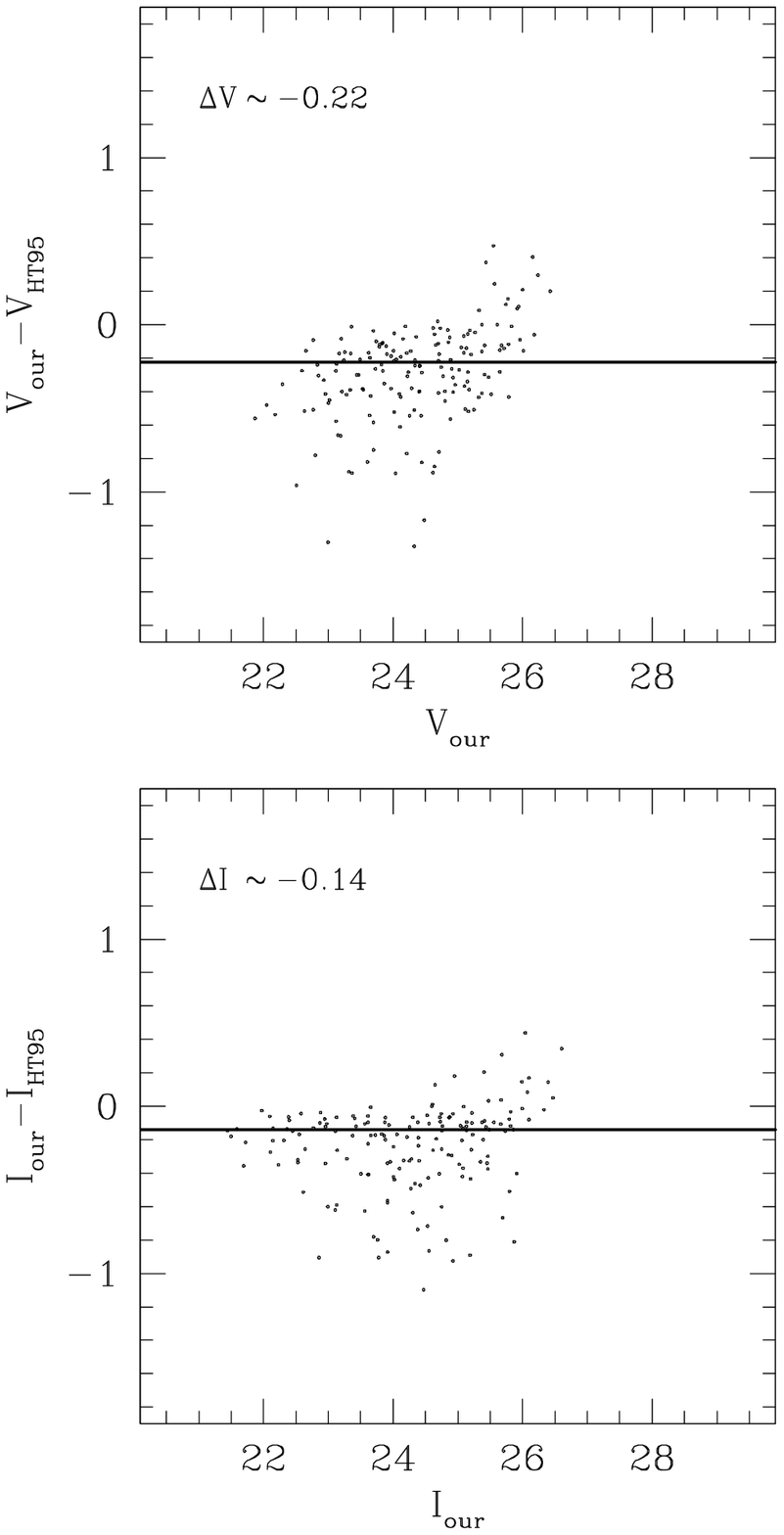}
%6

\clearpage

\epsscale{0.5}
%\plotone{VIcmdPC.eps}
\plotone{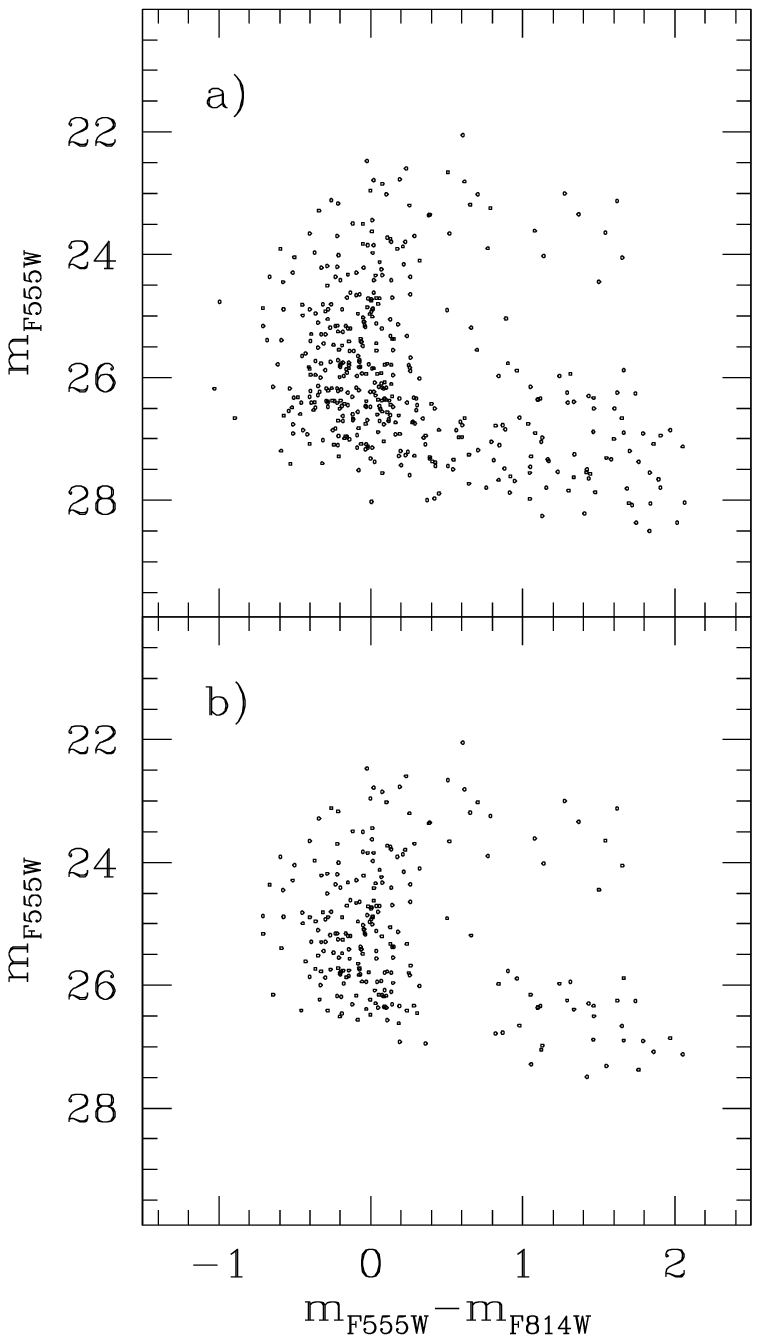}
%7

\clearpage

%\plotone{BVcmdPC.eps}
\plotone{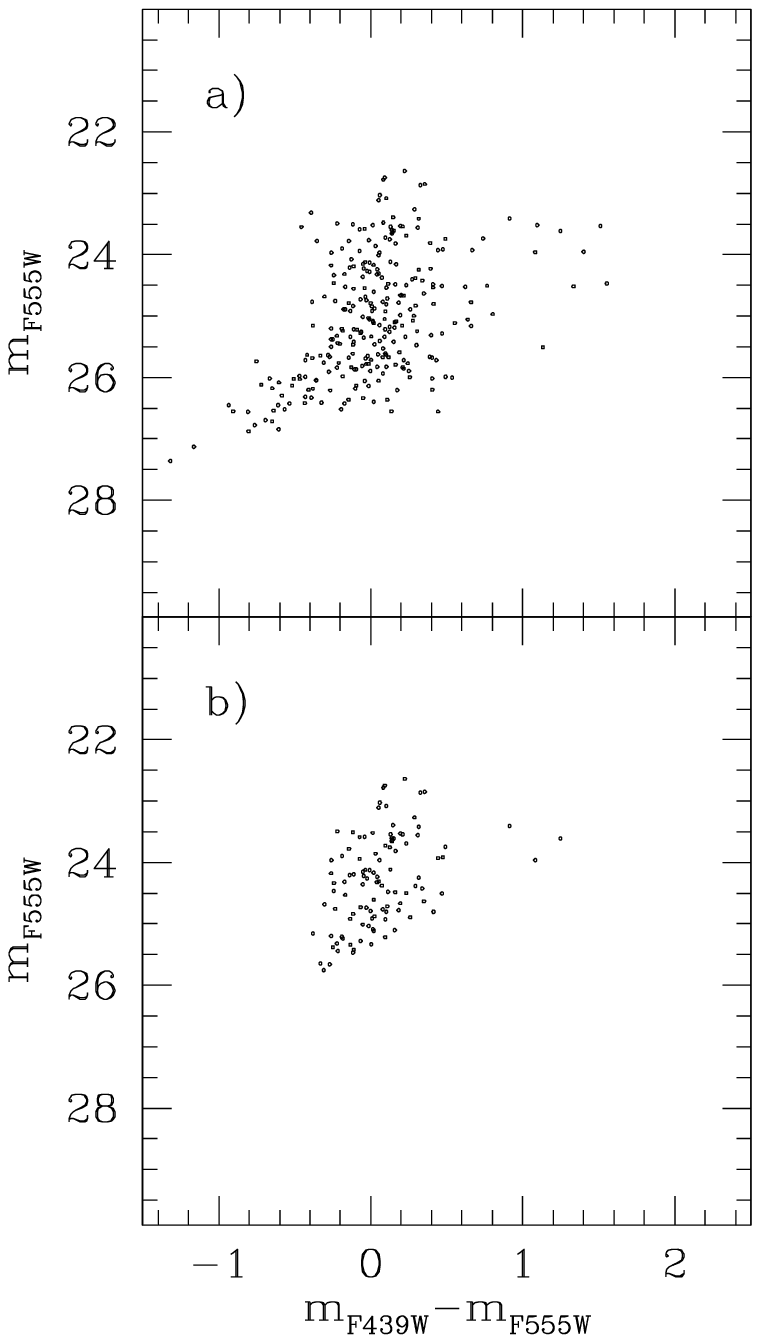}
%8

\clearpage

\epsscale{0.55}
%\plotone{tracksPC.eps}
\plotone{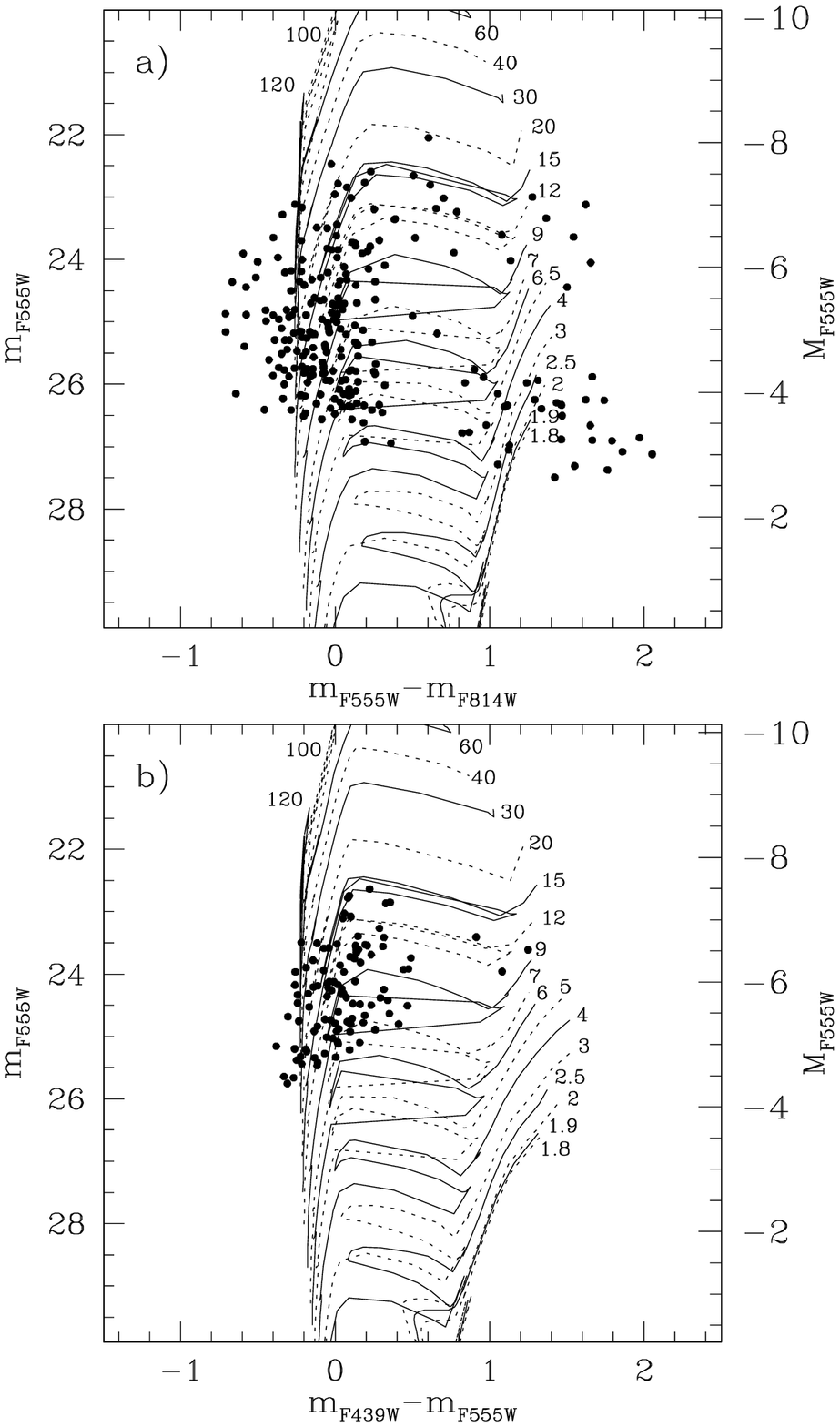}
%9

\clearpage

\epsscale{0.55}
%\plotone{PCLFs.eps}
\plotone{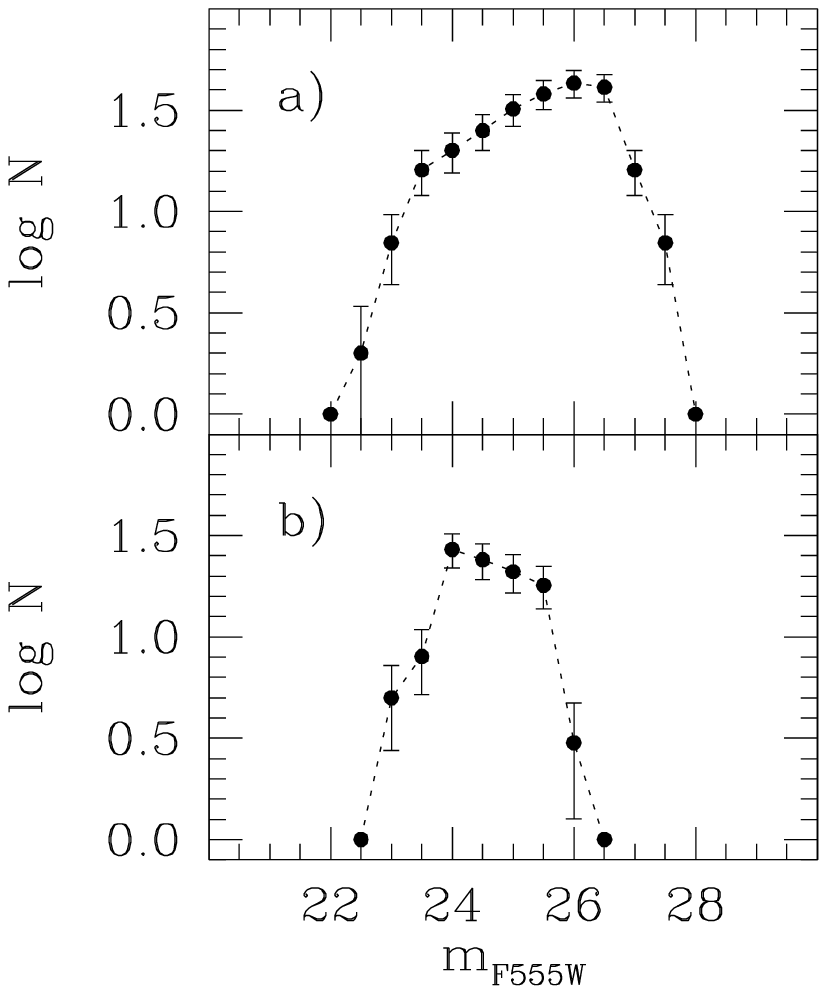}
%10

\clearpage

\epsscale{0.5}
%\plotone{BVcmdWF3.eps}
\plotone{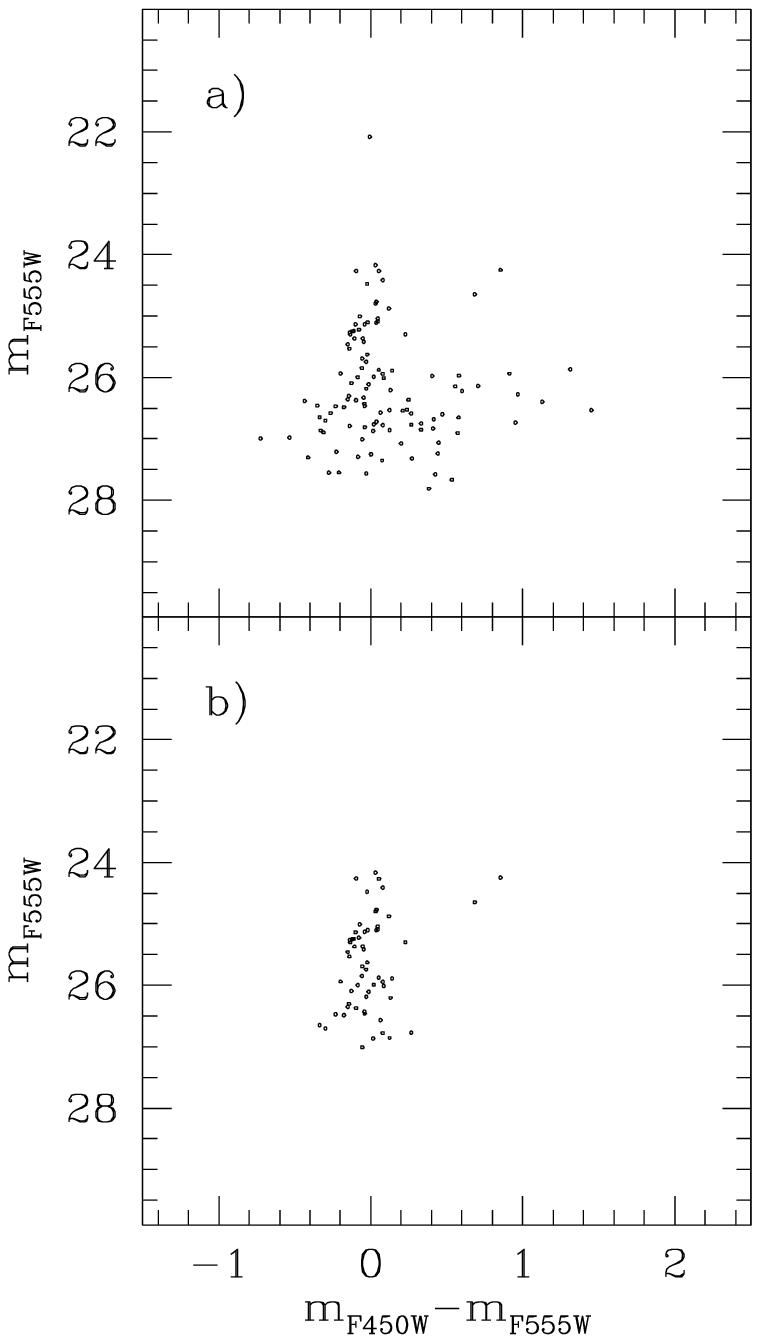}
%11

\clearpage

\epsscale{0.55}
%\plotone{WF3LF.eps}
\plotone{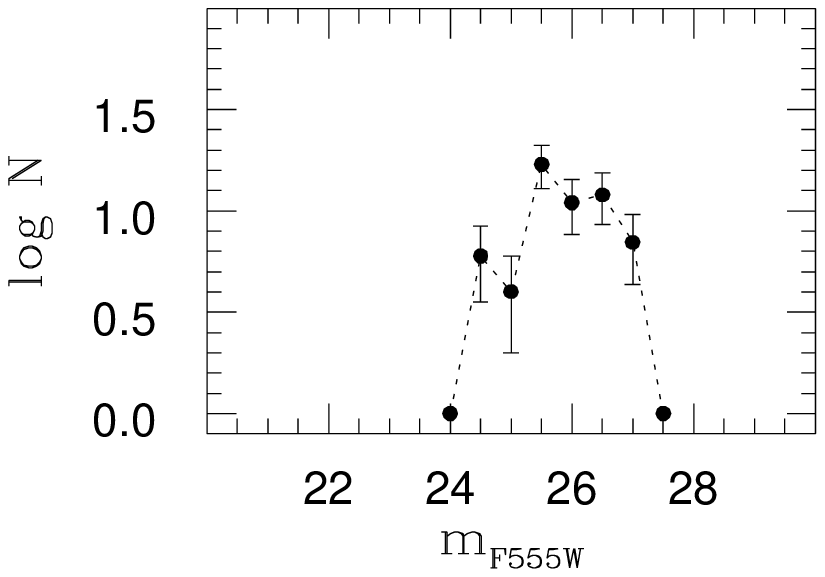}
%12

\clearpage

%\plotone{tracksWF3.eps}
\plotone{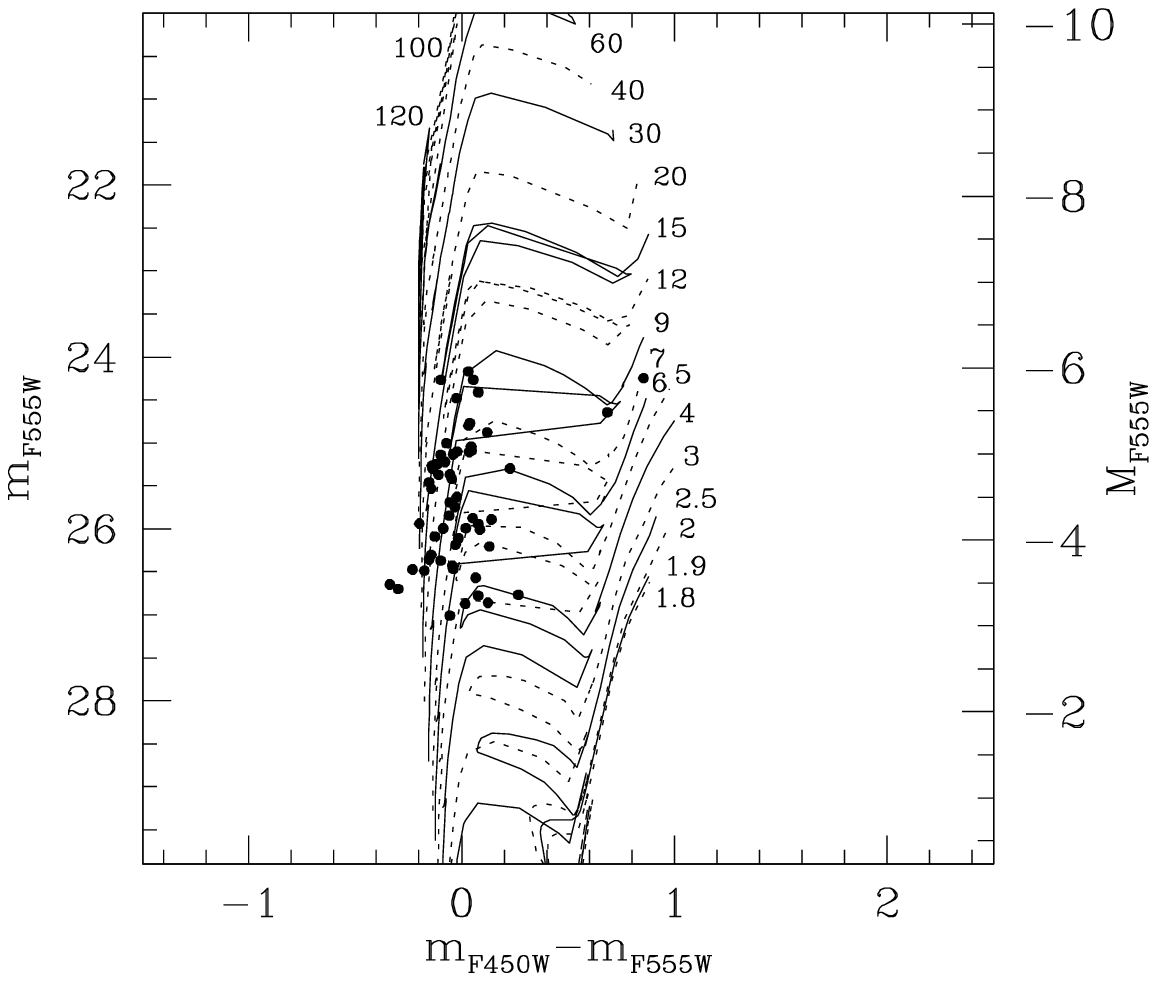}
%13

\clearpage

\epsscale{1.0}
%\plotone{sim1fin.eps}
\plotone{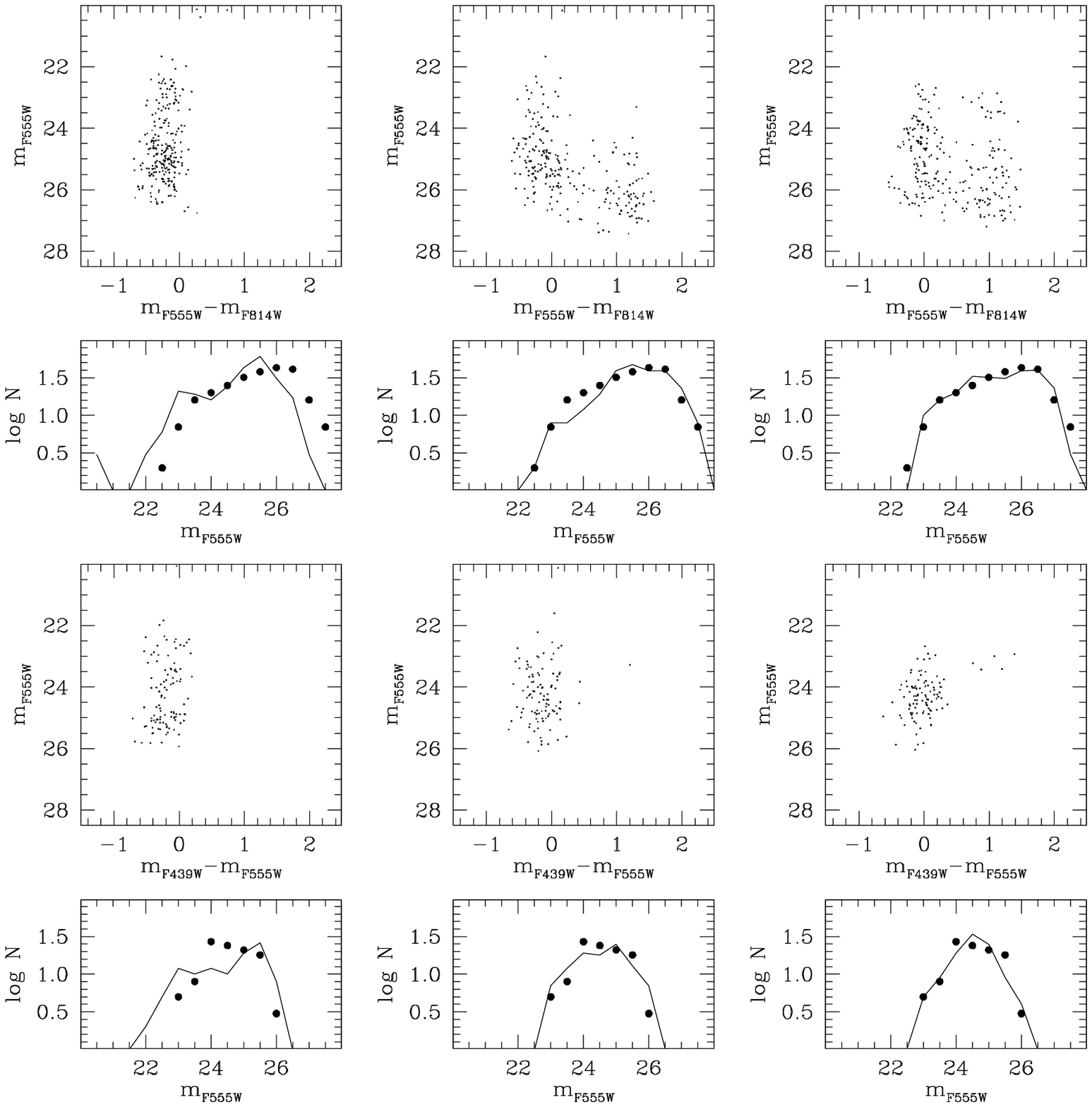}
%14

\clearpage

%\plotone{sim3nn.eps}
\plotone{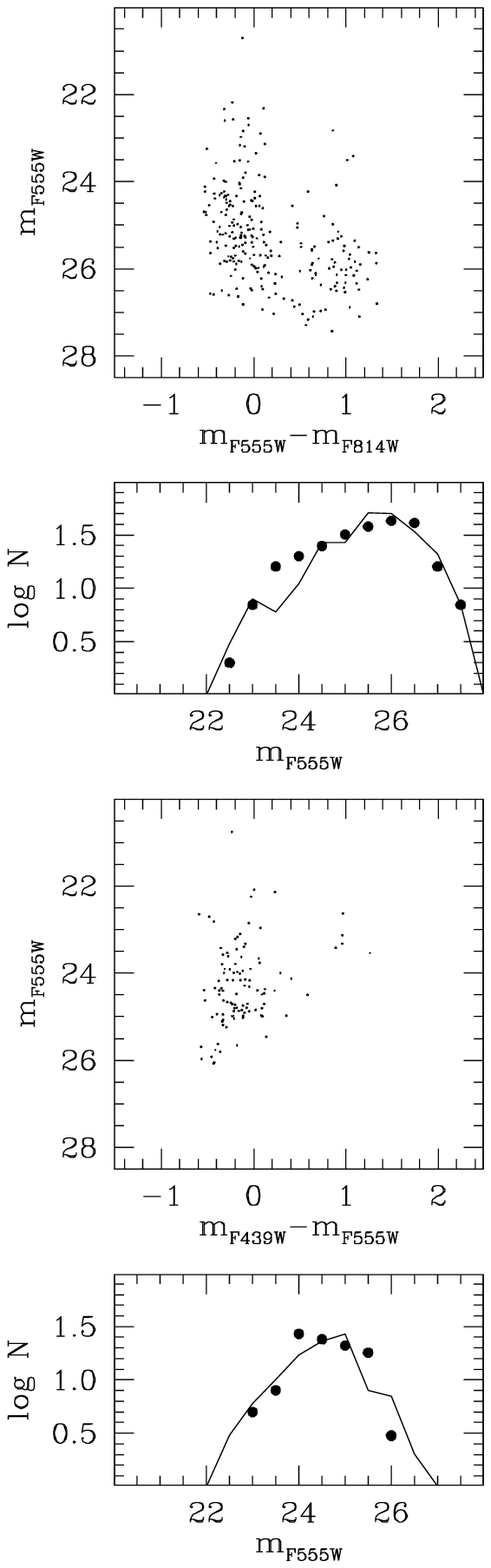}
%15

\clearpage

\epsscale{0.55}
%\plotone{tracksPC1.eps}
\plotone{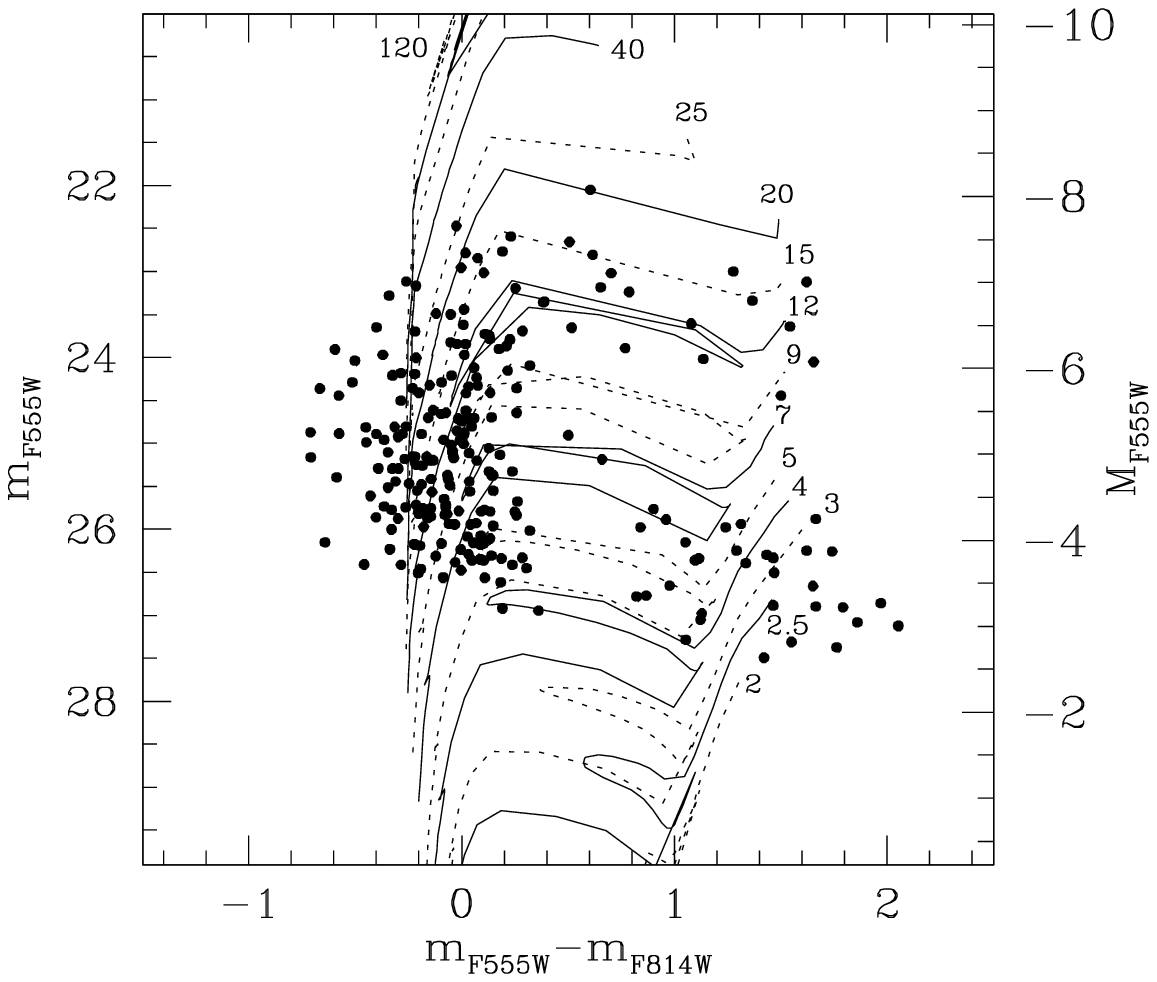}
%16

\clearpage

\epsscale{1.0}
%\plotone{sim2fin.eps}
\plotone{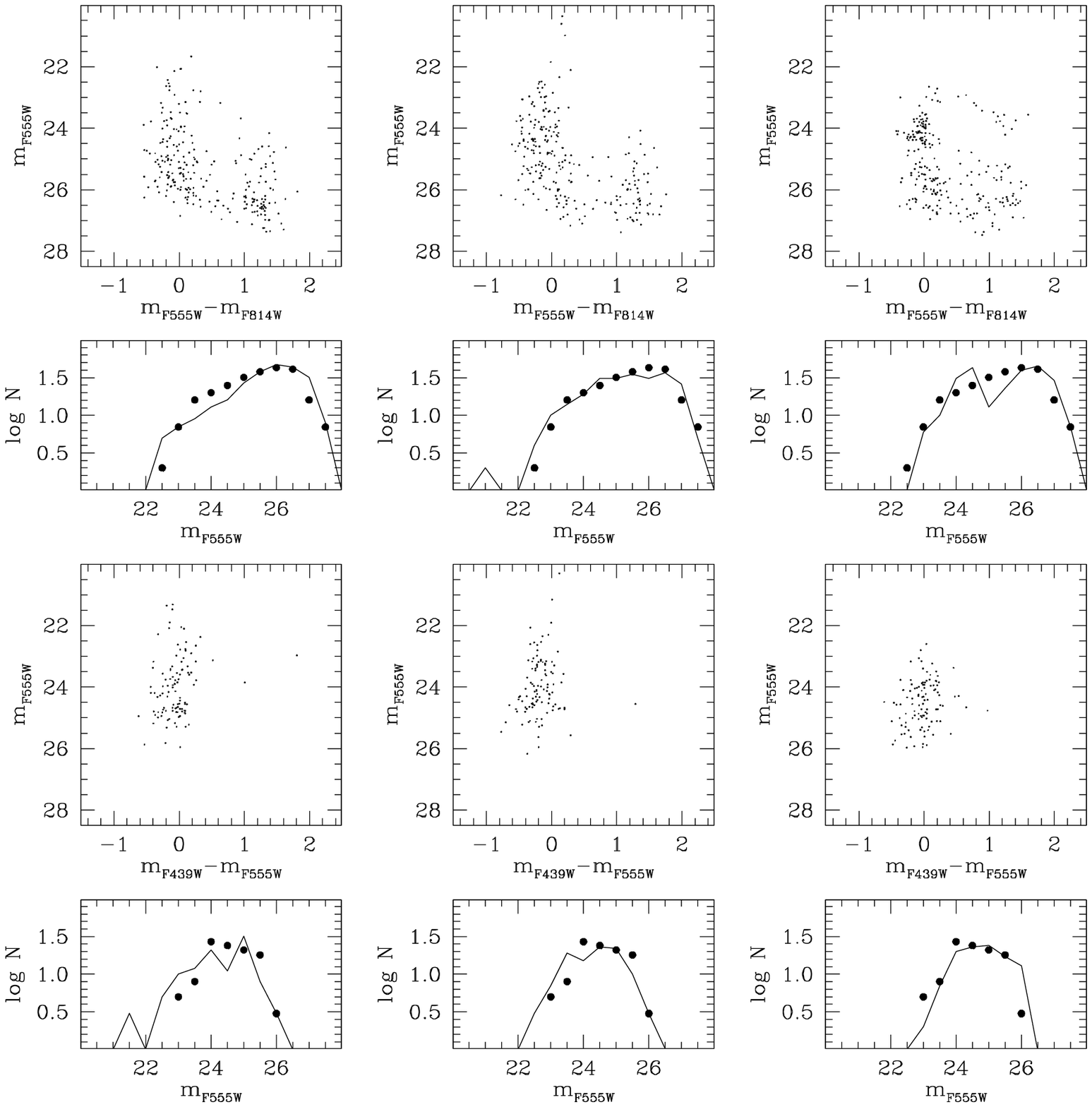}
%17

\clearpage

%\plotone{simsec1fin.eps}
\plotone{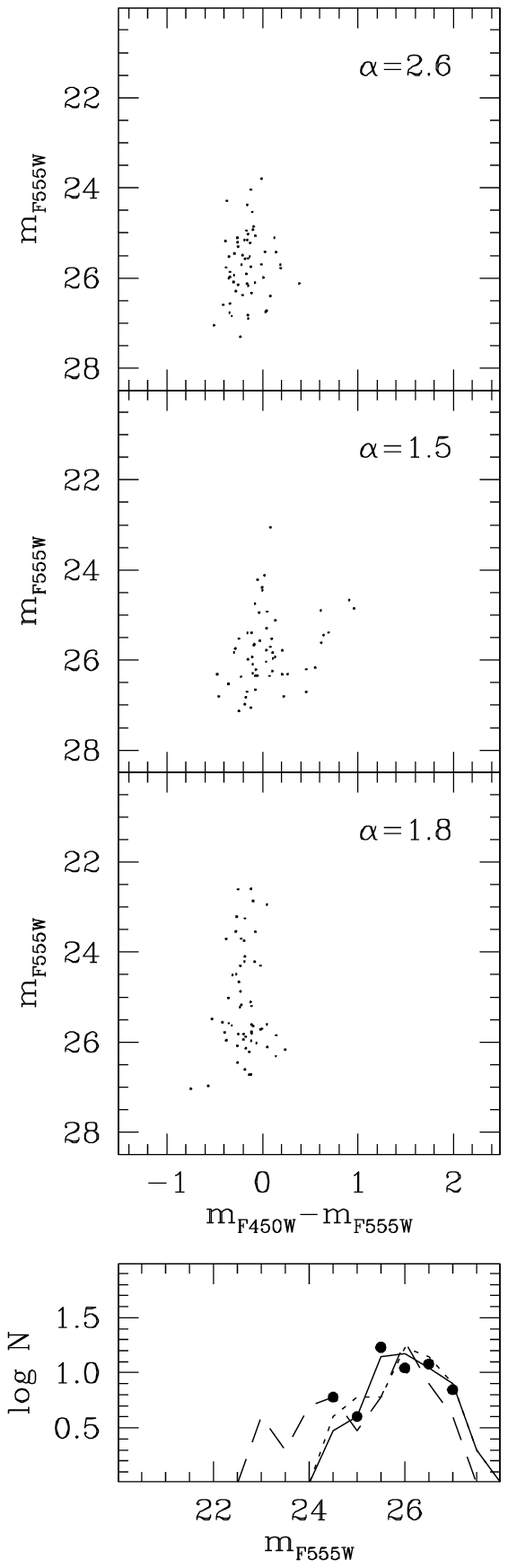}
%18

\end{document}